\documentclass[10pt,aps,prd,notitlepage,preprintnumbers]{revtex4-1}
\usepackage{graphicx}
\usepackage{slashed}
\usepackage{xcolor}
\usepackage{amsmath}
\usepackage{amssymb}
\usepackage{amsfonts}
\usepackage[utf8]{inputenc}
\usepackage[bookmarks=false,colorlinks, linkcolor=green]{hyperref}
\newcommand{\ep}{\epsilon}
\newcommand{\la}{\lambda}
\newcommand{\de}{\delta}
\newcommand{\ga}{\gamma}
\newcommand{\al}{\alpha}
\newcommand{\be}{\beta}
\newcommand{\sig}{\sigma}
\newcommand{\s}[1]{\slashed{#1}}

\newcommand{\no}{\notag\\}

\newcommand{\tx}[1]{\text{#1}}
\begin{document}
\title{Back-to-back heavy quark pair production in Semi-inclusive DIS}
\author{Guang-Peng \surname{Zhang}}
\affiliation{Department of modern physics, University of Science and Technology of China, Hefei, Anhui 230026, China}

\begin{abstract}
The one-loop correction to heavy quark pair back-to-back production in unpolarized semi-inclusive deep inelastic scattering is given in this work in the framework of transverse momentum dependent(TMD) factorization. Both unpolarized and linearly polarized TMD gluon distribution functions are taken into account. A subtraction method based on diagram expansion is used to get finite hard coefficients. It is found the soft and collinear divergences of one-loop amplitude is proportional to tree level ones and can be expressed through several basic scalar triangle and bubble integrals. The subtraction of these divergences is spin independent. Beyond tree level an additional soft factor related to final heavy quark pair must be added into the factorization formula. This soft factor affects the azimuthal angle distribution of virtual photon in a nonperturbative way. Integrating over virtual photon azimuthal angle we construct three weighted cross sections, which depend on only three additional integrated soft factors. These weighted cross sections can be used to extract linearly polarized gluon distribution function. In addition, lepton azimuthal angle is unintegrated in this work, which provides more observables. All hard coefficients relevant to lepton and virtual photon azimuthal angle distributions are given at one-loop level.
\end{abstract}

\maketitle
\section{Introduction}
Gluon transverse momentum dependent(TMD) distribution functions take important information about the hadron structure. In high energy region, gluon as a parton usually plays more important role than quark, since the momentum fraction of parton is always very small. Thus, precise knowledge of gluon distribution functions is desired to make precise prediction of the cross section. Since gluon is spin-1, it can have different polarizations in a hadron. When the transverse momentum of the gluon in a hadron is integrated over, there is only one gluon distribution function possible for unpolarized hadron, which is the usual collinear gluon parton distribution function(PDF). In such integrated PDF the gluon is unpolarized. But when the transverse momentum of gluon is unintegrated, the transverse momentum of gluon can couple with the spin of gluon to give a description of polarized distribution of gluon in a hadron\cite{Mulders:2000sh}. For unpolarized hadron, there exists two such transverse momentum dependent gluon distribution functions: one is denoted as $G(x,p_\perp^2)$, which reflects the distribution of unpolarized gluon in a hadron, with longitudinal momentum fraction $x$ and transverse momentum $p_\perp$, the other is $H^\perp(x,p_\perp^2)$, which reflects the distribution of linearly polarized gluon in the hadron. So far, the second gluon distribution function, i.e., $H^\perp(x,p_\perp^2)$ has already aroused much interest in current researches, such as the effect of $H^\perp$ on higgs boson production at LHC\cite{Boer:2011kf,Boer:2013fca,Boer:2014tka,Sun:2011iw,Echevarria:2015uaa}. One can see e.g.,\cite{Angeles-Martinez:2015sea,Boer:2015uqa} for a review on linearly polarized gluon distribution function. Various schemes are proposed to extract this function in literature. For hadron reaction, pair photon production\cite{Qiu:2011ai}, photon-quarkonium\cite{Dunnen:2014eta} or quarkonium-quarkonium\cite{Zhang:2014vmh} associated production are proposed to extract this function from a $\cos(2\phi)$ azimuthal angle distribution in these reactions. For hadron reaction, the problem is the potential breaking of TMD factorization as pointed out in \cite{Collins:2007nk},\cite{Rogers:2010dm}. All these proposed schemes require the detected final states are colorless(for quarkonium the heavy quark pair from hard interaction is in a color singlet). Contrast to hadron reaction, semi-inclusive deep inelastic scattering(SIDIS) for heavy quark pair or di-jet production is expect to have an exact TMD factorization, and can be used to extract some definite information about gluon TMD distribution functions. \cite{Boer:2010zf} have examined the effect of $H^\perp$ in heavy quark pair and di-jet production. For heavy quark pair production, the study in \cite{Boer:2010zf} is at tree level or leading order of $\al_s$ with a simplified TMD formula. At this order the TMD formula just contains gluon TMD distribution functions as nonperturbative quantities. To higher order of $\al_s$, the soft radiation from final heavy quark pair will introduce an azimuthal angle dependent soft factor into TMD formula and affect the azimuthal angle distribution of virtual photon,i.e., $\phi_q$, which is used to extract $H^\perp$ in \cite{Boer:2010zf}. Because of the azimuthal angle dependent soft factor, a complete description of $\phi_q$ distribution in the cross section is impossible. Instead we construct three azimuthal angle weighted cross sections, which depend on only three integrated soft factors. These weighted cross sections may help to extract $H^\perp$.
In \cite{Zhu:2013yxa}, the TMD factorization for this process is examined at one-loop level, with $H^\perp$ not taken into account. It is found all collinear and soft divergences can be absorbed into gluon distribution function and a soft factor with azimuthal angle integrated. But part of the finite correction of hard coefficients is absent in \cite{Zhu:2013yxa}, and the study is confined to $G(x,p_\perp^2)$. In this paper, we want to study the azimuthal angle effect introduced by the soft factor mentioned above. We will examine the factorization formula for both $G(x,p_\perp)$ and $H^\perp(x,p_\perp)$ based on diagram expansion method presented in \cite{Ma:2013aca}. This method is different from the method in \cite{Zhu:2013yxa}, which uses a single gluon to replace initial hadron. The method based on diagram expansion enables us to obtain the finite hard coefficients for various parton distributions in a systematic way. In addition, we keep lepton azimuthal angle unintegrated in this work, which can provide more observables. The structure of this paper is as follows: in Sect.II we illustrate kinematics for heavy quark pair production in SIDIS; in Sect.III, tree level factorization and resulting angular distributions are discussed, including lepton azimuthal angle distributions; in Sect.IV, the factorization formula is examined at one-loop level and the hard coefficients are calculated. In this section, the soft factors are also constructed and virtual one-loop corrections to soft factors and distribution functions are calculated; in Sect.V, the effect of soft factor is discussed and three weighted cross sections are constructed to extract $H^\perp$; Sect.VI is our summary.

\section{Kinematics}
We consider the scattering of electron and hadron
\begin{align}
e(l)+h_A(P_A)\rightarrow e(l')+Q(k_1)+\bar{Q}(k_2)+X,
\end{align}
where $X$ represents undetected hadrons\cite{Diehl:2005pc}. At order of $\al_{em}^2$, this process is dominated by the exchange of a virtual photon between electron and hadron. The momentum of the virtual photon is $q^\mu=l^{\mu}-{l'}^\mu$. In perturbative region,
$Q^2=-q^2\gg \Lambda_{QCD}^2$. $Q,\bar{Q}$ with momenta $k_1,k_2$ are heavy quark and anti-quark produced in the hard collision. In the center of mass(c.m.) frame of virtual photon $\ga^*$ and initial hadron $h_A$, we demand $Q$ and $\bar{Q}$ are nearly back-to-back.

For convenience we define
\begin{align}
K^\mu=k_1^\mu+k_2^\mu,\ R^\mu=\frac{1}{2}(k_1^\mu-k_2^\mu).
\end{align}
Our requirement for the final quarks then becomes
\begin{align}
R_T^\mu\sim Q,\ K_T^\mu\ll Q,
\end{align}
where $R_T$ and $K_T$ are the transverse components of $R$ and $K$, respectively. In the c.m. frame of $\ga^*$ and $h_A$, the transverse vector is relative to $Z$-axis. Here $\vec{P}_A$ is along $+Z$-axis, and $\vec{q}$ is along $-Z$-axis.

By considering only the contribution of virtual photon, the cross section we want to study can be written as
\begin{align}
d\sig=\frac{1}{2s}\frac{e^4 Q_q^2}{Q^4}(2\pi)^4 \frac{d^3 l'}{(2\pi)^3 2{l'}^0}
\frac{d^3 k_1}{(2\pi)^3 2E_1}\frac{d^3 k_2}{(2\pi)^3 2E_2}L^{\mu\nu}W_{\mu\nu},
\end{align}
where the leptonic and hadronic tensors are
\begin{align}
L^{\mu\nu}=& 2(l^\mu {l'}^\nu+l^\nu {l'}^\mu-g^{\mu\nu}l\cdot l')\doteq 4l^\mu l^\nu +q^2 g^{\mu\nu};\no
W^{\mu\nu}=&\sum_X\langle P_A s|j^\nu|Q\bar{Q}X\rangle \langle Q\bar{Q}X |j^\mu(0)|P_A s\rangle\de^4(P_A+q-k_1-k_2-P_X).
\end{align}
In leptonic tensor we have used QED gauge invariance $q^\mu W_{\mu\nu}=0$ to eliminate all $q^\mu$ and $q^\nu$ in leptonic tensor. This will simplify our calculation.

Define
\begin{align}
x_B=\frac{Q^2}{2P_A\cdot q},\hspace{0.5cm}y=\frac{P_A\cdot q}{P_A\cdot l},
\hspace{0.5cm}z_1=\frac{P_A\cdot k_1}{P_A\cdot q},
\hspace{0.5cm}z_2=\frac{P_A\cdot k_2}{P_A\cdot q},\hspace{0.5cm}s=(P_A+l)^2.
\end{align}
The phase space integration measure can be written as
\begin{align}
\frac{d^3 l'}{2{l'}^0}=\frac{ys}{4}dx_B dy d\psi_l,\hspace{1cm}
\frac{d^3 k_1}{2E_1}\frac{d^3 k_2}{2E_2}=\frac{1}{4} dy_1 dy_2 d^2 K_T d^2 R_T,
\end{align}
where $\psi_l$ is the azimuthal angle between ${l'}_T$ and $R_T$; $y_{1,2}$ are the rapidities of final quark and anti-quark, respectively.

Then, the differential cross section becomes
\begin{align}
\frac{d\sig}{dx_B dy d\psi_l dy_1 dy_2 d^2K_T d^2 R_T}=
\frac{y \al_{em}^2 Q_q^2}{64\pi^3 Q^4}L^{\mu\nu}W_{\mu\nu},
\end{align}
where $Q_q$ is the electric charge of heavy quark in unit of the charge of electron.
All information about the cross section now is contained in the hadronic tensor $W^{\mu\nu}$. In next sections we will focus on the factorization of hadronic tensor.

$\ga^* N$ frame(c.m. frame of $\ga^*$ and $h_A$) is useful to describe the cross section, but it is not very convenient for the calculation, especially for the factorization of $W^{\mu\nu}$, as we discussed later. More convenient frame is the hadron frame defined as the c.m. frame of final heavy quark and antiquark. In this frame, $\vec{P}_A$ is still along $+Z$-axis, but virtual photon now has a small transverse momentum $q_\perp$. We use light-cone coordinate representation in this paper. In this representation any vector $a^\mu$ is written as $(a^+,a^-,a_\perp^\mu)$, where the transverse components are relative to $Z$-axis and denoted by $a_\perp$; $\pm$-components are defined by two light-like vectors $n^\mu$ and $\bar{n}^\mu$ with $n\cdot\bar{n}=1$, i.e., $a^+= n\cdot a$, $a^-=\bar{n}\cdot a$. The decomposition of relevant momenta is like
\begin{align}
P_A^\mu= P_A^+ \bar{n}^\mu,\ K^\mu=K^+ \bar{n}^\mu+K^- n^\mu,\ q^\mu=q^+ \bar{n}^\mu+q^- n^\mu +q_\perp^\mu.
\end{align}
The following two tensors are useful to project transverse components of a vector:
\begin{align}
g_\perp^{\mu\nu}=g^{\mu\nu}-n^\mu\bar{n}^\nu-n^\nu\bar{n}^\mu,\ \ \ep_\perp^{\mu\nu}=\ep^{\mu\nu-+}=\ep^{\mu\nu\rho\tau}\bar{n}_{\rho}n_\tau,
\end{align}
and our convention for $\ep$-tensor is $\ep^{0123}=1$ so that $\ep_\perp^{12}=1$. For any four-vector $a^\mu$ its transverse component is $a_\perp^\mu=g_\perp^{\mu\nu}a_\nu$.

Given these two frames, $q_\perp$ in hadron frame and $K_T$ in $\ga^*N$ frame can be connected to each other. In $\ga^* N$ frame,
\begin{align}
K_T^\mu=K^\mu -\al P_A^\mu -\be q^\mu.
\end{align}
Hence, the transverse component of $K_T$ in the hadron frame is
\begin{align}
(K_T)_\perp^\mu=-\be q_\perp^\mu=-(z_1+z_2)q_\perp^\mu.
\end{align}
On the other hand,
\begin{align}
(K_T)_\perp^\mu=g_\perp^{\mu\nu}K_{T\nu}=K_T^\mu-P_A^\mu\frac{K_T^2}{P_A\cdot K}.
\end{align}
Combining these two equations we have
\begin{align}
K_T^\mu=-(z_1+z_2)q_\perp^\mu + P_A^\mu\frac{K_T^2}{P_A\cdot K}.
\end{align}
This is an exact result. In the region we are considering, $K_T$ is a small quantity, then the second term with $K_T^2$ can be ignored at leading power level.

\begin{figure}
\includegraphics[scale=0.5]{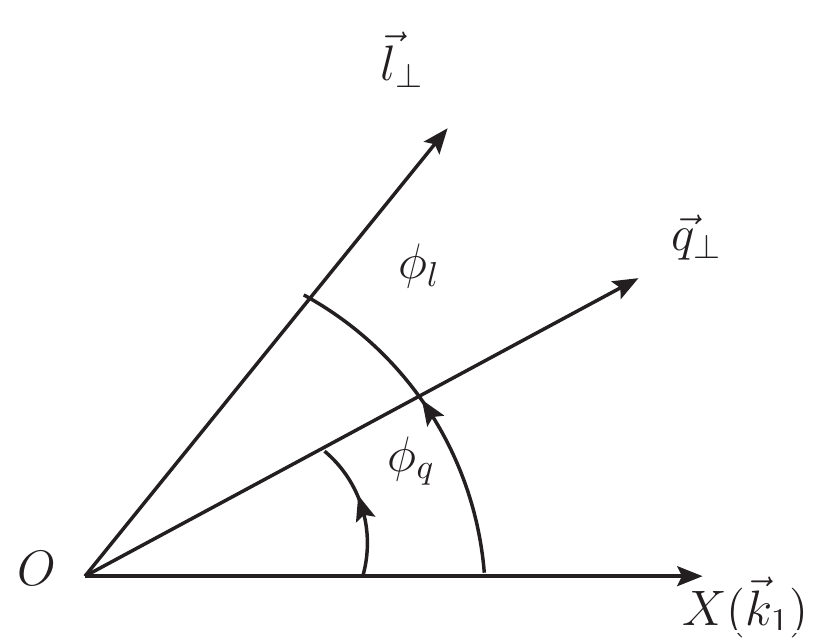}
\caption{Azimuthal angles of virtual photon and initial lepton in the transverse plane of hadron frame. }
\label{fig:azimuthal-ql}
\end{figure}

In this work all calculations will be performed in hadron frame. To define azimuthal angles on the transverse plane we assign $\vec{k}_{1\perp}$ along $+X$-axis, and $Y$-axis is defined by $\ep_\perp$-tensor, that is,
\begin{align}
X^\mu=\frac{k_{1\perp}^\mu}{|\vec{k}_{1\perp}|},\ Y^\mu=\ep_\perp^{\mu\nu}X_\nu.
\end{align}
Then the azimuthal angles of virtual photon and initial lepton, $\phi_q$ and $\phi_l$, are defined as the angles of $q_\perp$ and $l_\perp$ relative to $X$-axis, respectively, as shown in Fig.\ref{fig:azimuthal-ql}. At leading power of $q_\perp$ expansion, $\phi_l$ is equal to $\psi_l$, which is defined in $\ga^*N$ frame.
With these two vectors $X^\mu$ and $Y^\mu$ the transverse metric and anti-symmetric tensor can be written into another form
\begin{align}
g_\perp^{\mu\nu}=-X^\mu X^\nu -Y^\mu Y^\nu,\ \ep_\perp^{\mu\nu}=X^\mu Y^\nu-X^\nu Y^\mu.
\end{align}
Since $n,\bar{n},X,Y$ form a complete basis in four-dimension space, the leptonic tensor can be expressed through these four vectors. Equivalently, one can choose
$P_A,K,X,Y$ as basis. The advantage of this basis is the calculation can be done in a covariant way.

The complete basis for symmetric rank-2 tensor is
\begin{align}
A_i^{\mu\nu}=&\left\{\tilde{n}^\mu \tilde{n}^\nu, \tilde{n}^\mu X^\nu+\tilde{n}^\nu X^\mu,\ \tilde{n}^\mu Y^\nu+\tilde{n}^\nu Y^\mu,\
X^\mu X^\nu+Y^\mu Y^\nu,\ X^\mu X^\nu-Y^\mu Y^\nu,\ X^\mu Y^\nu+X^\nu Y^\mu \right\},\hspace{0.5cm}i=1,\cdots,6.
\label{eq:Ai}
\end{align}
With this basis leptonic tensor in hadron frame is expressed as
\begin{align}
L^{\mu\nu}=& Q^2\frac{4(1-y)}{y^2} \tilde{n}^\mu \tilde{n}^\nu + Q^2\frac{1+(1-y)^2}{y^2}(X^\mu X^\nu+Y^\mu Y^\nu)\no
&+ 2Q^2\frac{\sqrt{1-y}(y-2)}{y^2}\cos\phi_l (\tilde{n}^\mu X^\nu + \tilde{n}^\nu X^\mu)+2Q^2\frac{\sqrt{1-y}(y-2)}{y^2}\sin\phi_l (\tilde{n}^\mu Y^\nu + \tilde{n}^\nu Y^\mu)\no
&+ 2Q^2\frac{1-y}{y^2}\cos(2\phi_l)(X^\mu X^\nu-Y^\mu Y^\nu)+2Q^2\frac{1-y}{y^2}\sin(2\phi_l)(X^\mu Y^\nu+X^\nu Y^\mu),
\label{eq:leptonic-tensor}
\end{align}
where
\begin{align}
\tilde{n}^\mu=\frac{1}{\sqrt{2\al_k P_A\cdot K+\al_k^2 K^2}}(P_A+\al_k K)^\mu,\ \al_k=-\frac{P_A\cdot q}{K\cdot q}.
\label{eq:ntilde}
\end{align}
In the decomposition we have considered the constraint of QED gauge invariance, that is, $P_A^\mu$ and $K^\mu$ must be combined into $\tilde{n}^\mu$ to ensure $q_\mu A_i^{\mu\nu}=0$. Since $q\cdot X\sim q\cdot Y\sim \mathcal{O}(q_\perp)$, QED gauge invariance is preserved at leading power of $q_\perp$. One can check that $\tilde{n}^\mu$ can have another representation like
\begin{align}
\tilde{n}^\mu=\frac{1}{Q}(q+2 x_B P_A)^\mu+\mathcal{O}(q_\perp).
\end{align}
This representation simplifies our calculation dramatically. Since hadronic tensor does not depend on $\phi_l$, this decomposition exhibits all possible lepton azimuthal angle distributions for unpolarized lepton beam.

\section{Tree level azimuthal angle dependence}
In hadron frame when $q_\perp$ is small, $W^{\mu\nu}$ can have a TMD factorization formula at the leading power of $q_\perp$ expansion. At tree level the formula reads
\begin{align}
W^{\mu\nu}=&\frac{x_B}{x Q^2 M_p(N_c^2-1)}\de(1-z_1-z_2)H^{\mu\nu}_{\al\be}\int d^2 p_\perp \de^2(p_\perp+q_\perp)\Phi^{\al\be}(x,p_\perp)
+\mathcal{O}(\frac{q_\perp}{Q},\frac{q_\perp}{k_{1\perp}}),
\label{eq:hadronic-tree}
\end{align}
where the gluon TMDPDF is\cite{Mulders:2000sh}
\begin{align}
\Phi^{\al\be}(x,p_\perp)=& \left(\frac{p^+}{2M_p}\right)^{-1}\int\frac{d\xi^- d^2\xi_\perp}{(2\pi)^3}e^{ix \xi^-P_A^++i\xi_\perp\cdot p_\perp}
\langle P_A|G_{a\perp}^{+\be}(0)G_{a\perp}^{+\al}(0^+,\xi^-,\xi_\perp)|P_A\rangle\no
=& -g_\perp^{\al\be}G(x,p_\perp^2)+\frac{2p_\perp^\al p_\perp^\be-g_\perp^{\al\be}p_\perp^2}{2M_p^2}H^\perp(x,p_\perp^2),
\end{align}
We work in Feynman gauge $\partial^\mu G_\mu=0$ in this work. The gauge links in $\Phi^{\al\be}$ are suppressed for simplicity. There is a color summation in the definition of gluon TMDPDF. Therefore, there is a color average in the hard part.
In formula eq.(\ref{eq:hadronic-tree}), the average factor $1/(N_c^2-1)$ has been extracted, so, the hard part $H^{\mu\nu}_{\al\be}$ in eq.(\ref{eq:hadronic-tree}) contains a summation over color.

\begin{figure}
\includegraphics[scale=0.5]{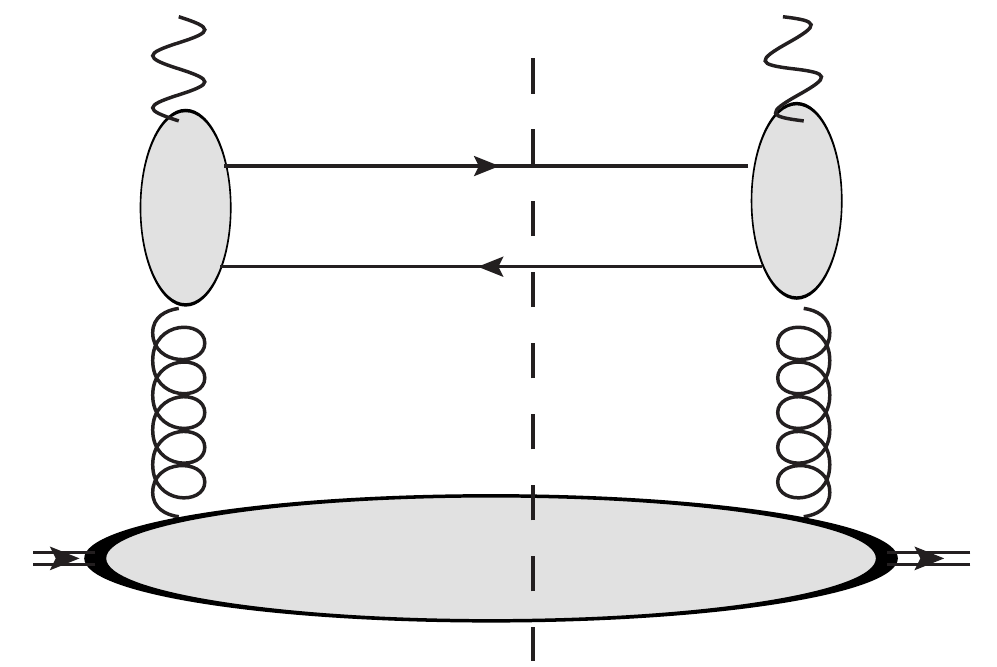}
\caption{TMD factorization for heavy quark pair production in DIS. The two central bubbles represent hard subprocess, while lower bubble represents jet part of initial hadron.}
\label{fig:QQ-prod}
\end{figure}

The derivation of this factorization formula for hadronic tensor has been given in \cite{Pisano:2013cya} in detail. Under high energy limit or $Q^2\to \infty$, the general structure of the interaction factorizes into the form shown in Fig.\ref{fig:QQ-prod}, where the two central bubbles represent hard interaction, in which all propagators are far off-shell, and the lower bubble represents the jet part of initial hadron, in which all propagators are collinear to $P_A$. All possible soft interactions are ignored in Fig.\ref{fig:QQ-prod}. These soft interactions do not appear at the leading order of $\al_s$. We will discuss them in one-loop correction.
For this process, the hard scales are $Q$ and $k_{1\perp}$, which are taken to be the same order in this paper. Under the limit $q_\perp\ll Q, k_{1\perp}$, the above factorization formula is obtained from the expansion in $\la\simeq q_\perp/Q,\ q_\perp/k_{1\perp}$. Leading power contribution of this process is given by collinear partons, that is, the momentum of initial gluon satisfies $p^\mu=(p^+,p^-,p_\perp)\sim Q(1,\la^2,\la)$. According to this scaling law $p_\perp$, $q_\perp$ are of the same order, and then the delta function $\de^2(p_\perp+q_\perp)$ itself is already of leading power. Therefore, both $p_\perp$ and $q_\perp$ can be ignored in $H^{\mu\nu}_{\al\be}$. This means $H^{\mu\nu}_{\al\be}$ are on-shell amplitudes. This fact ensures QED gauge invariance of hadronic tensor.

\begin{figure}
\includegraphics[scale=0.5]{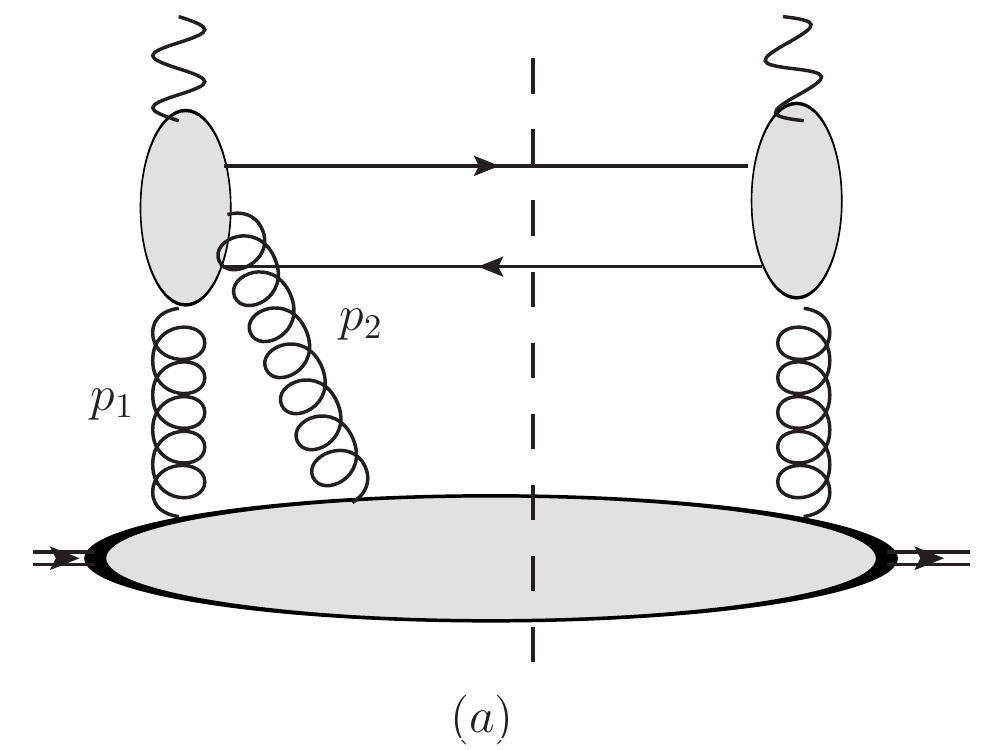}\hspace{2cm}
\includegraphics[scale=0.5]{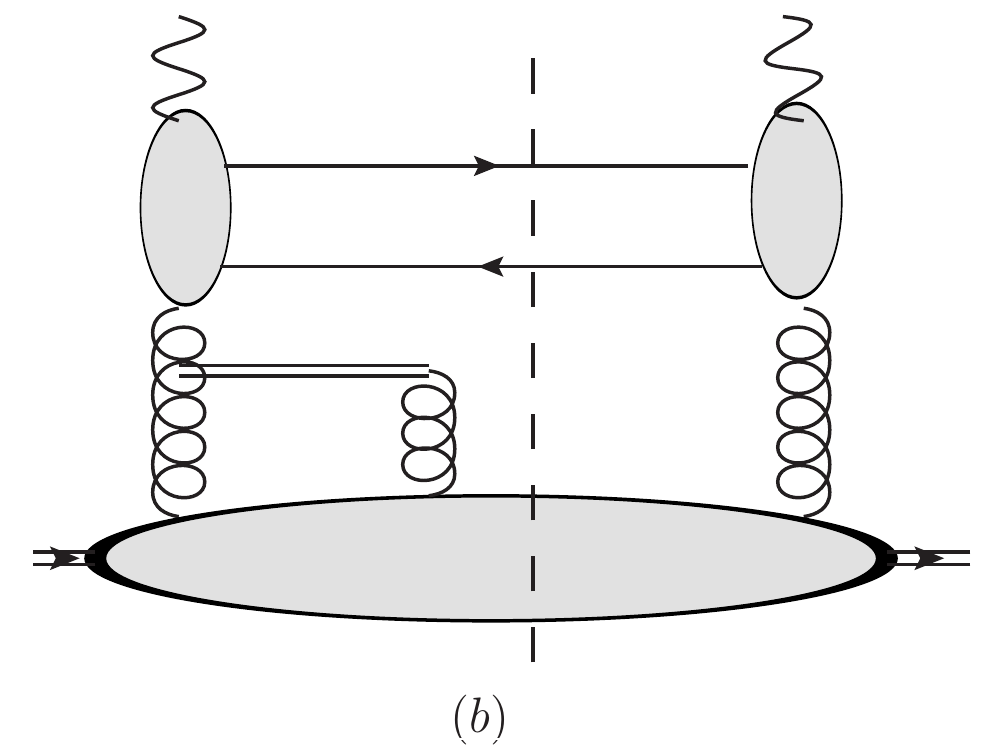}
\caption{Scattering with two collinear gluons in initial state. The two initial gluons cannot couple into one gluon through tri-gluon vertex.}
\label{fig:2gluon}
\end{figure}

For factorization in Feynman gauge gluon distributions may have a problem concerning super-leading power contribution, which appears when the gluon in
Fig.\ref{fig:QQ-prod} is longitudinally polarized, i.e., $G^+$.
Such a gluon will give a $1/\Lambda$ enhancement compared to the leading power contribution we stated above. Note that the delta function
in the hard part, $\de^2(p_\perp+q_\perp)$, should not be expanded, then super-leading power contribution is given by the longitudinal gluon with $p_\perp=0$. For the one-gluon case considered here, such a contribution from longitudinally gluon vanishes due to Ward identity.
But in Feynman gauge, there can be any number of longitudinal gluons connecting the central bubble and lower bubble, whose contribution is not power suppressed. For the case with two gluons connecting the central bubble and lower bubble as shown in Fig.\ref{fig:2gluon}(a), \cite{Collins:2008sg} has given an explicit calculation to show the super-leading power contribution is absent even when the transverse momentum of the parton is preserved. In addition, at leading power one of the two gluons becomes gluon field strength tensor, the other is absorbed into gauge link as shown in Fig.\ref{fig:2gluon}(b). With this conclusion we will simply take the gluon in Fig.\ref{fig:QQ-prod} as transversely polarized even at one-loop level, and will consider only the one-gluon case in our calculation. This causes no problem about the hard coefficients at least at one-loop level.

Because $p_\perp$ is set to zero, there is only one independent transverse momentum $k_{1\perp}$ in $H^{\mu\nu}_{\al\be}$. Then previous vector basis for $L^{\mu\nu}$ can also be applied to $H^{\mu\nu}_{\al\be}$. That is,
\begin{align}
H^{\mu\nu}_{\al\be}=\sum_{i,j}H_{ij}A_i^{\mu\nu}B^j_{\al\be},
\label{eq:hard-coeff}
\end{align}
with $A_i^{\mu\nu}$ given by eq.(\ref{eq:Ai}) and
\begin{align}
B_j^{\al\be}=&\left\{X^\al X^\be+Y^\al Y^\be,\ X^\al X^\be-Y^\al Y^\be,\ X^\al Y^\be+X^\be Y^\al \right\},\hspace{0.5cm}j=1,2,3.
\end{align}
Due to P-parity conservation the number of $Y$ vector in $A_i B_j$ must be even. So, there are only following 10 rather than 18 nontrivial projected hard coefficients:
\begin{align}
H_{ij}=\{H_{11},H_{12},H_{21},H_{22},H_{33},H_{41},H_{42},H_{51},H_{52},H_{63}\}.
\end{align}
With these projected hard coefficients, the angular distributions on $\phi_q$ and $\phi_l$ can be obtained as
\begin{align}
&\frac{d\sig}{d\psi_l dy dx_B dy_1 dy_2 d^2 K_T d^2 R_T }
\left[\frac{Q_q^2 \al_{em}^2 x_B y\de(1-z_1-z_2)}{16\pi^3 Q^4 x M_p(N_c^2-1)}\right]^{-1}\no
=&\frac{2(1-y)}{y^2}\left(H_{11}\langle w_1 G\rangle -\cos(2\phi_q) H_{12}\langle w_2 H^\perp\rangle \right)
+\frac{1+(1-y)^2}{y^2}\left(H_{41}\langle w_1 G\rangle -\cos(2\phi_q) H_{42}\langle w_2 H^\perp\rangle \right)\no
&-\frac{2\sqrt{1-y}(y-2)}{y^2}\cos\phi_l\left(H_{21}\langle w_1 G\rangle -\cos(2\phi_q) H_{22}\langle w_2 H^\perp\rangle \right)
+\frac{2\sqrt{1-y}(y-2)}{y^2}\sin\phi_l \sin(2\phi_q) H_{33}\langle w_2 H^\perp\rangle \no
&+\frac{2(1-y)}{y^2}\cos{2\phi_l}\left(H_{51}\langle w_1 G\rangle -\cos(2\phi_q) H_{52}\langle w_2 H^\perp\rangle \right)
-\frac{2(1-y)}{y^2}\sin{2\phi_l}\sin(2\phi_q) H_{63}\langle w_2 H^\perp\rangle.
\end{align}
where
\begin{align}
\langle w(p_\perp, q_\perp) f(x,p_\perp^2)\rangle \equiv \int d^2 p_\perp \de^2(p_\perp+q_\perp) w(p_\perp, q_\perp) f(x,p_\perp^2),
\end{align}
and
\begin{align}
w_1(p_\perp, q_\perp)=1,\ w_2(p_\perp, q_\perp)=\frac{2(p_\perp\cdot q_\perp)^2-p_\perp^2 q_\perp^2}{2M_p^2 q_\perp^2}.
\end{align}

\begin{figure}
\includegraphics[width=0.23\textwidth]{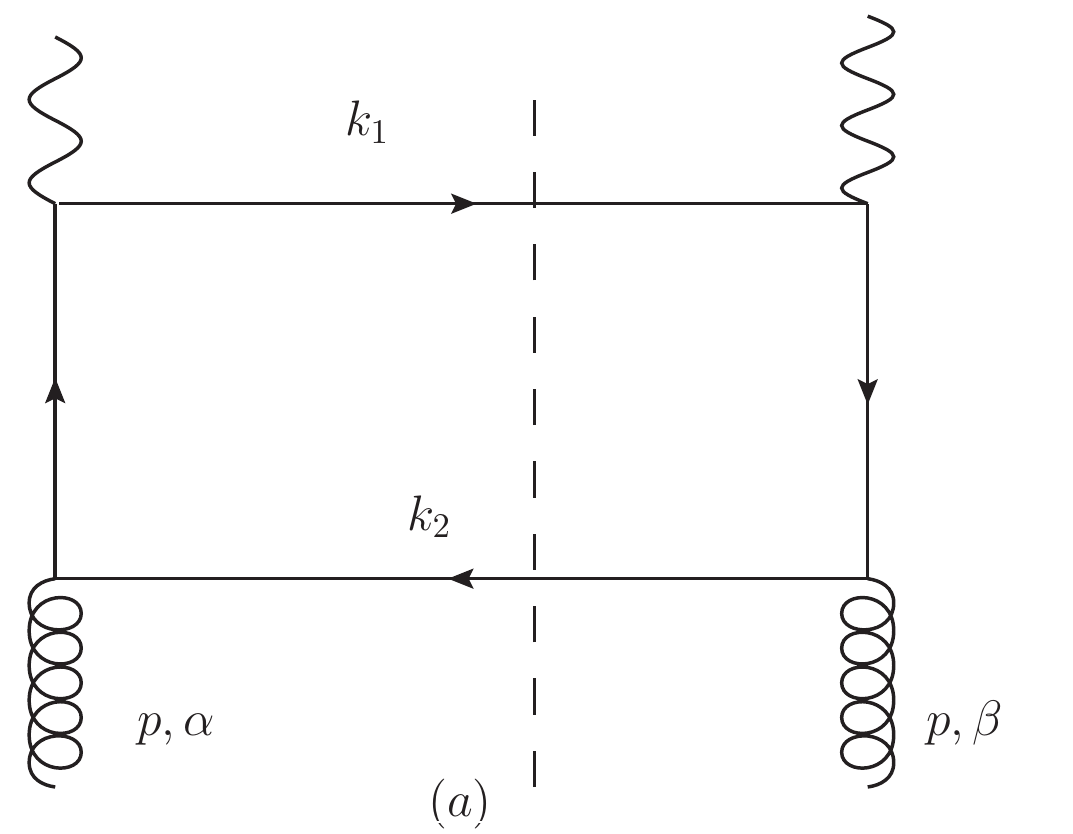}
\includegraphics[width=0.2\textwidth]{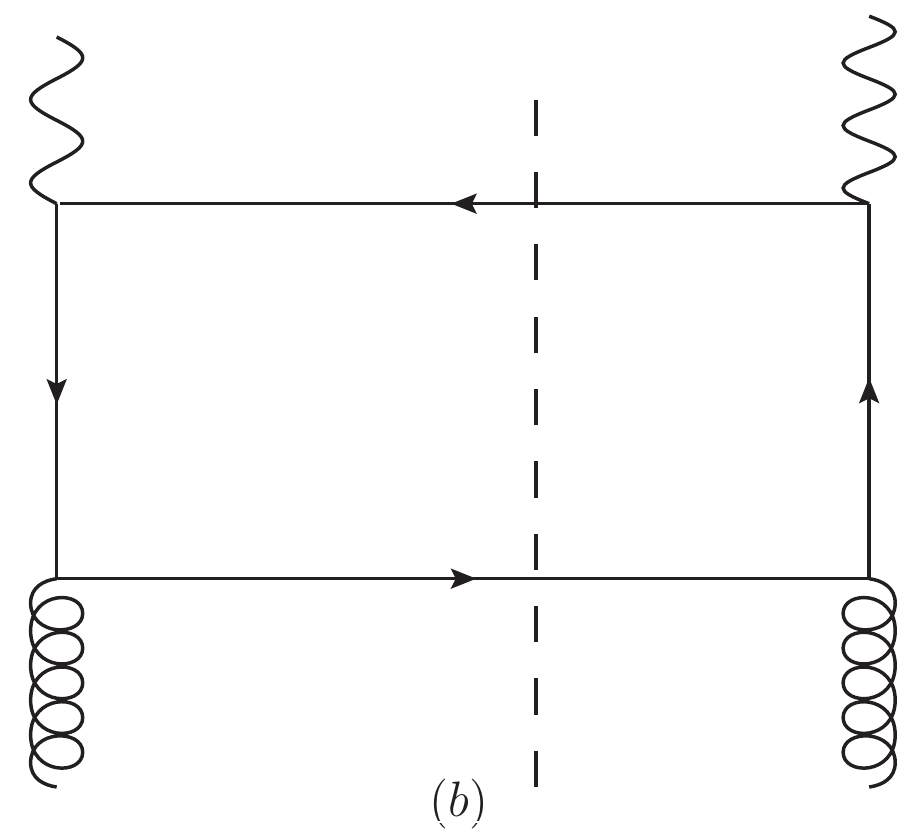}
\includegraphics[width=0.2\textwidth]{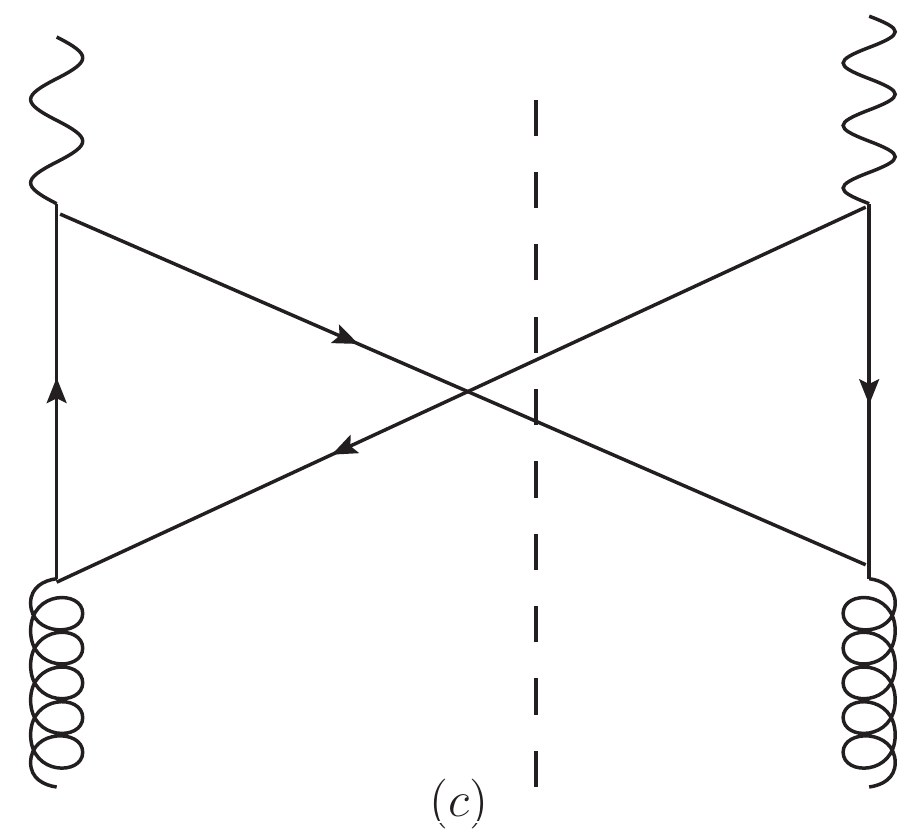}
\includegraphics[width=0.2\textwidth]{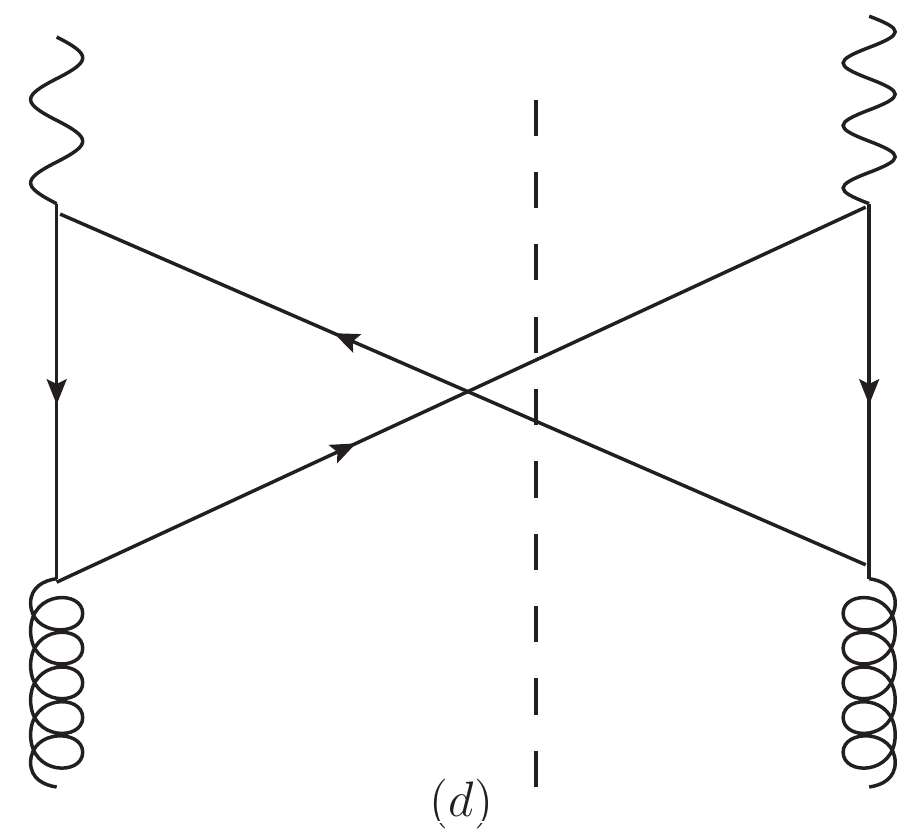}
\caption{Tree diagrams for the subprocess $\ga^* g\to Q\bar{Q}$. }
\label{fig:tree}
\end{figure}

The tree level hard coefficients can be obtained by replacing the central bubbles in Fig.\ref{fig:QQ-prod} with the diagrams in Fig.\ref{fig:tree}. For the subprocess
$\ga^*+g\to Q\bar{Q}$, there are three independent variables:
\begin{align}
s_1=2k_1\cdot k_2,\ t_1=-2p\cdot k_1,\ u_1=-2p\cdot k_2,
\end{align}
where $p^\mu=x P_A^\mu$ is the momentum of initial gluon.
Another independent parameter we choose in our calculation is quark mass $m$.
For convenience we define
\begin{align}
\tilde{H}_{ij}=&\frac{1}{g_s^2 N_cC_F} H_{ij}.
\end{align}
After some simplifications the results are
\begin{align}
\tilde{H}_{11}=&\tilde{H}_{12}=\frac{16 \left(2 m^2+s_1+t_1+u_1\right)
   \left(m^2
   \left(t_1^2+u_1^2\right)-s_1 t_1
   u_1\right)}{t_1 u_1
   \left(t_1+u_1\right){}^2}=\frac{16 Q^2 |R_\perp|^2}{t_1 u_1 },\no
\tilde{H}_{21}=&\tilde{H}_{22}=Q|R_\perp|\frac{4 \left(t_1-u_1\right) \left(t_1
   u_1 \left(2 s_1+t_1+u_1\right)-2 m^2
   \left(t_1^2+u_1^2\right)\right)}{t_1^
   2 u_1^2 \left(t_1+u_1\right)},\no
\tilde{H}_{33}=&Q|R_\perp| \frac{4(t_1-u_1)}{t_1 u_1},\no
\tilde{H}_{41}=&\frac{2}{t_1^2 u_1^2
   \left(t_1+u_1\right){}^2} \left(-8 m^4 \left(t_1^3 u_1+2
   t_1^2 u_1^2+t_1
   u_1^3+t_1^4+u_1^4\right)\right.\no
   &-2 m^2
   \left(s_1 \left(-4 t_1^3 u_1-2 t_1^2
   u_1^2-4 t_1
   u_1^3+t_1^4+u_1^4\right)+\left(t_1+u_
   1\right)
   \left(t_1^2+u_1^2\right){}^2\right)\no
   &\left.+t
   _1 u_1 \left(t_1^2+u_1^2\right)
   \left(2 s_1 \left(t_1+u_1\right)
   +2
   s_1^2+\left(t_1+u_1\right){}^2\right)
   \right),\no
\tilde{H}_{42}=&-\frac{4 \left(m^2
   \left(t_1^2+u_1^2\right)-s_1 t_1
   u_1\right) \left(4 m^2 \left(t_1
   u_1+t_1^2+u_1^2\right)+\left(t_1^2+u_
   1^2\right)
   \left(s_1+t_1+u_1\right)\right)}{t_1^
   2 u_1^2 \left(t_1+u_1\right){}^2},\no
\tilde{H}_{51}=&\frac{8 \left(t_1 u_1
   \left(s_1+t_1+u_1\right)-m^2
   \left(t_1^2+u_1^2\right)\right)
   \left(m^2
   \left(t_1^2+u_1^2\right)-s_1 t_1
   u_1\right)}{t_1^2 u_1^2
   \left(t_1+u_1\right){}^2},\no
\tilde{H}_{52}=&-\frac{4 \left(2 m^4
   \left(t_1^2+u_1^2\right){}^2-2 m^2
   t_1 u_1 \left(t_1^2+u_1^2\right)
   \left(2 s_1+t_1+u_1\right)+t_1^2
   u_1^2 \left(2 s_1
   \left(t_1+u_1\right)+2
   s_1^2+\left(t_1+u_1\right){}^2\right)
   \right)}{t_1^2 u_1^2
   \left(t_1+u_1\right){}^2},\no
\tilde{H}_{63}=&\frac{4 \left(2 m^2
   \left(t_1^2+u_1^2\right)-t_1 u_1
   \left(2
   s_1+t_1+u_1\right)\right)}{t_1 u_1
   \left(t_1+u_1\right)},
\label{eq:hard-coeff-tree}
\end{align}
where
\begin{align}
Q^2=-q^2=-(2m^2+s_1+t_1+u_1),\ R_\perp^2=-|R_\perp|^2=k_{1\perp}^2=\frac{m^2(t_1^2+u_1^2)-s_1 t_1 u_1}{(t_1+u_1)^2}.
\end{align}
From C-parity conservation $H_{21,22,33}$ are anti-symmetric, while other ones are symmetric in $t_1$ and $u_1$. Our result satisfies this symmetry. The momentum fraction $x$ can be obtained from $s_1+t_1+u_1=-2m^2-Q^2$, $t_1=-2p\cdot k_1=-2x P_A\cdot k_1$ and $u_1=-2p\cdot k_2=-2x P_A\cdot k_2$. The explicit value is
\begin{align}
x= x_B\frac{s_1+2m^2+Q^2}{Q^2(z_1+z_2)}=x_B\frac{s_1+2m^2+Q^2}{Q^2},
\label{eq:x}
\end{align}
in which all variables can be measured in experiment and in the last equality we have used $z_1+z_2=1$.

\section{One-loop correction}
For TMD factorization here, the relative transverse momentum of final heavy quark and antiquark is fixed. The soft divergence from virtual correction cannot be cancelled by real correction, since the phase space integration is incomplete for real correction. Usually, a soft factor is introduced to absorb such soft divergences. The operator form of the soft factor can be obtained by using eikonal approximation for soft gluons emitted by final heavy quark pair and by initial gluon, see \cite{Kidonakis:1997gm} for example. The procedure is standard and the heavy quark soft factor is
\begin{align}
S^{Q}(b_\perp)=&\frac{1}{Tr(T^c T^c)}\langle 0|\hat{U}^\dagger_{\tilde{v}}(b_\perp,-\infty)_{ae}Tr[U_{v_2}(\infty,b_\perp)T^e U^\dagger_{v_1}(\infty,b_\perp)
U_{v_1}(\infty,0)T^dU^\dagger_{v_2}(\infty,0)]_{ed}\hat{U}_{\tilde{v}}(0,-\infty)|0\rangle.
\label{eq:SQ}
\end{align}
The definition of gauge link is
\begin{align}
U_{v}(\infty,b_\perp)=&P \exp\left[-ig_s\int_0^\infty d\la v\cdot G(b_\perp+\la v)\right],\no
\hat{U}_v(b_\perp,-\infty)=&P \exp\left[-ig_s\int_{-\infty}^0 d\la v\cdot\hat{G}(b_\perp+\la v)\right],
\end{align}
with $v$ an arbitrary vector and $P$ the path-ordering product so that fields with smaller $\la$ are always put on right hand side of the fields with larger $\la$. $U_v$ and $\hat{U}_v$ are defined in fundamental and adjoint representations of color group, respectively. Correspondingly, $G^\mu=G^\mu_a T^a$ and $\hat{G}^\mu =G^\mu_a \hat{T}^a$ are gluon fields in fundamental and adjoint representations. Note that $\hat{T}^c_{ba}=-if^{cba}$ and $Tr(T^a T^b)=\de^{ab}/2$ in this work. With such definitions of the gauge link our covariant derivative is $D^\mu=\partial^\mu +ig_s G_a^\mu T^a$. The $Tr[\cdots]$ in eq.(\ref{eq:SQ}) acts on matrices in fundamental representation. Then, one can check the definition eq.(\ref{eq:SQ}) is color gauge invariant. At order of $\al_s^{0}$, $S^Q(b_\perp)$ is normalized to 1. The definition in eq.(\ref{eq:SQ}) is given in coordinate space. To get the definition in momentum space one should do a Fourier transformation, i.e.,
\begin{align}
S^Q(l_\perp)=\int \frac{d^2 b_\perp}{(2\pi)^2} e^{ib_\perp\cdot l_\perp}S^Q(b_\perp).
\end{align}
At order of $\al_s^0$, $S^Q(l_\perp)=\de^2(l_\perp)$.

The sources of the gauge links in $S^Q$ are clear: $U_{v_1}$ and $U_{v_2}$ are obtained from the coupling of soft gluon to on-shell heavy quark or anti-quark by using eikonal
approximation. Here $v_{1,2}=k_{1,2}/m$ are the four-velocities of heavy quark and anti-quark, respectively; $\hat{U}_{\tilde{v}}$ is extracted from the coupling of soft gluon to initial hadron or gluon. Here $\tilde{v}$ is collinear to the momentum of initial hadron. If $\tilde{v}^2=0$, $S^Q$ has a light-cone divergence, and thus is not well-defined. As a regulator $\tilde{v}$ is modified to be a little away from the light-cone direction but still with $\tilde{v}_\perp$ vanishing, i.e., $\tilde{v}^+\gg\tilde{v}^-$ and $\tilde{v}_\perp=0$.

Without the soft factor the tree level TMD formula
eq.(\ref{eq:hadronic-tree}) cannot be right. The correct one should be
\begin{align}
W^{\mu\nu}
=&\frac{x_B\de(1-z_1-z_2)}{x Q^2 M_p(N_c^2-1)}\int d^2 p_\perp d^2 l_{1\perp} d^2 l_{2\perp}\de^2(p_\perp+q_\perp-l_{1\perp}-l_{2\perp})H^{\mu\nu}_{\al\be}
\Phi^{\al\be}(x,p_\perp)
S^Q(l_{1\perp})\bar{S}(l_{2\perp}),
\label{eq:oneloop-formula}
\end{align}
where $H^{\mu\nu}_{\al\be}$ is the hard part.
$\bar{S}(l_\perp)$ appears in order to avoid the double counting of soft divergences, since the soft divergence in the correction to gluon TMDPDF $\Phi^{\al\be}$ is also contained in the correction to $S^Q$. Except that now the gauge link is defined in adjoint representation, $\bar{S}(l_\perp)$ is the same as that defined in of SIDIS\cite{Ji:2004wu}, i.e.,
\begin{align}
\bar{S}(l_\perp)=&\int \frac{d^2 b_\perp}{(2\pi)^2}e^{ib_\perp\cdot l_\perp}
\frac{N_c^2-1}{\langle 0|\hat{U}^\dagger_{\tilde{v}}(b_\perp,-\infty)_{ae}\hat{U}^\dagger_v(\infty,b_\perp)_{ed}
\hat{U}_v(\infty,0)_{dc}\hat{U}_{\tilde{v}}(0,-\infty)_{ca}|0\rangle}.
\end{align}
In this soft factor the vector $\tilde{v}$ has appeared in $S^Q$. Another vector appearing in the gauge links is $v^\mu=(v^+,v^-,0_\perp)$ with $v^-\gg v^+$. As stated before, the little offshellness of $v$ and $\tilde{v}$ is used to regularize light-cone singularity. Other regulartors for light-cone singularity have been given by \cite{Collins:2011zzd},\cite{GarciaEchevarria:2011rb}. The calculation procedure with these regulators is the same, so, we will not calculate the hard coefficients once more using the regulators in \cite{Collins:2011zzd},\cite{GarciaEchevarria:2011rb}.

The calculation of one-loop hard coefficients can be performed in the same way as \cite{Ma:2013aca}. One-loop hard coefficient $H^{(1)}_{finite}$ is given by
\begin{align}
\int_\perp H^{(1)}_{finite}\Phi_{\al\be}(x,p_\perp) S_Q(l_{1\perp})\bar{S}(l_{2\perp})
=&
\int_\perp H^{(1)}\Phi_{\al\be}(x,p_\perp) S_Q(l_{1\perp})\bar{S}(l_{2\perp}) -\int_\perp H^{(0)}\left[\Phi^{(1)}_{\al\be}(x,p_\perp) S_Q(l_{1\perp})\bar{S}(l_{2\perp})\right.\no
&\left.+\Phi_{\al\be}(x,p_\perp) S^{(1)}_Q(l_{1\perp})\bar{S}(l_{2\perp})
+\Phi_{\al\be}(x,p_\perp) S_Q(l_{1\perp})\bar{S}^{(1)}(l_{2\perp})
\right],\no
\int_\perp=&\int d^2 p_\perp d^2 l_{1\perp} d^2 l_{2\perp}
\de^2(p_\perp+q_\perp-l_{1\perp}-l_{2\perp}),
\label{eq:subtraction-scheme}
\end{align}
where $H^{(1)},\Phi^{(1)},S_Q^{(1)},\bar{S}^{(1)}$ represents the one-loop corrections to hard scattering part, gluon TMDPDF, and the two soft factors, respectively. One-loop integral in $H^{(1)}$ includes parton-like contribution\cite{Collins:2008sg}, for which the loop integral is collinear to $p$ or $P_A$. This parton-like part has been included in tree level result and should be subtracted to avoid calculating tree level diagrams by twice. For details of subtraction one can consult \cite{Collins:2008sg},\cite{Ma:2013aca},\cite{Ma:2015vpt}. At leading power one can show that all real corrections to hadronic tensor can be subtracted by the correction to gluon distribution and soft factors. If the final gluon is collinear to $P_A$, \cite{Collins:2008sg} has shown that its coupling to heavy quarks can be absorbed into gauge link in gluon distributions. If the final gluon is soft, one can use eikonal approximation to transform the coupling of soft gluon to collinear gluon or to heavy quarks into gauge links in heavy quark soft factor $S^Q$. The overlap between $\Phi^{\al\be}$ and $S^Q$ in soft region is subtracted by another soft factor $\bar{S}$.

Hence, only virtual corrections contribute to one-loop hard coefficients. For virtual correction, the delta function $\de^2(p_\perp+q_\perp-l_{1\perp}-l_{2\perp})$ has been of leading power. So, $p_\perp,q_\perp$ can be ignored in one-loop hard part $H^{(1)}$, and then $H^{(1)}$ is the product of on-shell amplitudes for subprocess
$\ga^*+p\to Q\bar{Q}$. It is in this way the QED gauge invariance is preserved. If the TMD factorization formula is correct the subtracted hard coefficients $H^{(1)}_{finite}$ must be free of any infrared(IR) divergence. It should be noted that in formula eq.(\ref{eq:oneloop-formula}) all gluon TMDPDF, $S^{Q}$ and $\bar{S}$ are renormalized quantities, i.e., the UV divergences in these functions are removed by $\overline{MS}$ scheme. Thus these nonperturbative quantities all contain a renormalization scale $\mu$. In this work, both UV and IR divergences are regularized in dimensional scheme with $D=4-\ep$. Specially, in our scheme only loop momentum is generalized to $D$-dimension space, all other momenta are defined in four-dimension space. This is the four-dimensional-helicity(FDH) scheme(see \cite{Bern:1995db} and references therein). This scheme is convenient for the tensor decomposition of hard part as done in eq.(\ref{eq:hard-coeff}). Next we will first calculate the virtual correction to gluon TMDPDF and to the two soft factors and then present the structure of one-loop virtual correction to hadronic tensor. In the last subsection we present the explicit result of hard coefficients after subtraction.

\subsection{Virtual correction to nonperturbative quantities}
\begin{figure}
\begin{minipage}{0.45\textwidth}
\includegraphics[scale=0.5]{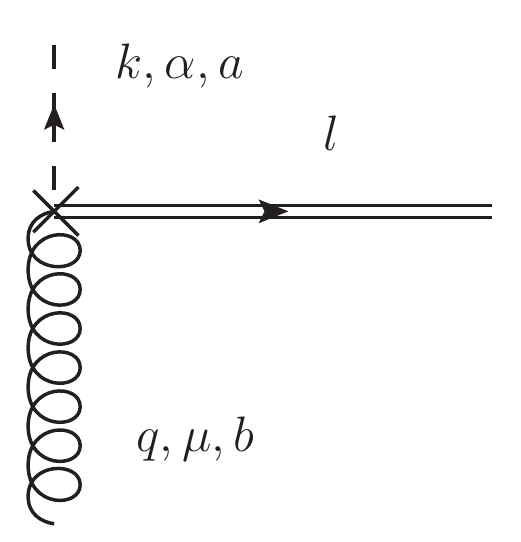}
\end{minipage}
\begin{minipage}{0.45\textwidth}
\begin{align}
-i\de_{ab}(n\cdot k g_\perp^{\al\mu}-n^\mu q_\perp^\al)\notag
\end{align}
\end{minipage}
\caption{The special vertex for the coupling of gluon field strength tensor and gauge link, where $q=k+l$.}
\label{fig:Phi-special}
\end{figure}

The complete gluon TMDPDF with gauge links is
\begin{align}
\Phi^{\al\be}(x,p_\perp)=& \left(\frac{p^+}{2M_p}\right)^{-1}\int\frac{d\xi^- d^2\xi_\perp}{(2\pi)^3}e^{-ix \xi^-P_A^+-i\xi_\perp\cdot p_\perp}
\langle P_A|G_{b\perp}^{+\be}(\xi)\hat{U}^\dagger_v(\infty,\xi)_{bc}
\hat{U}_v(\infty,0)_{ca}G_{a\perp}^{+\al}(0)|P_A\rangle_{\xi^+=0}.
\end{align}
\begin{figure}
\includegraphics[scale=0.5]{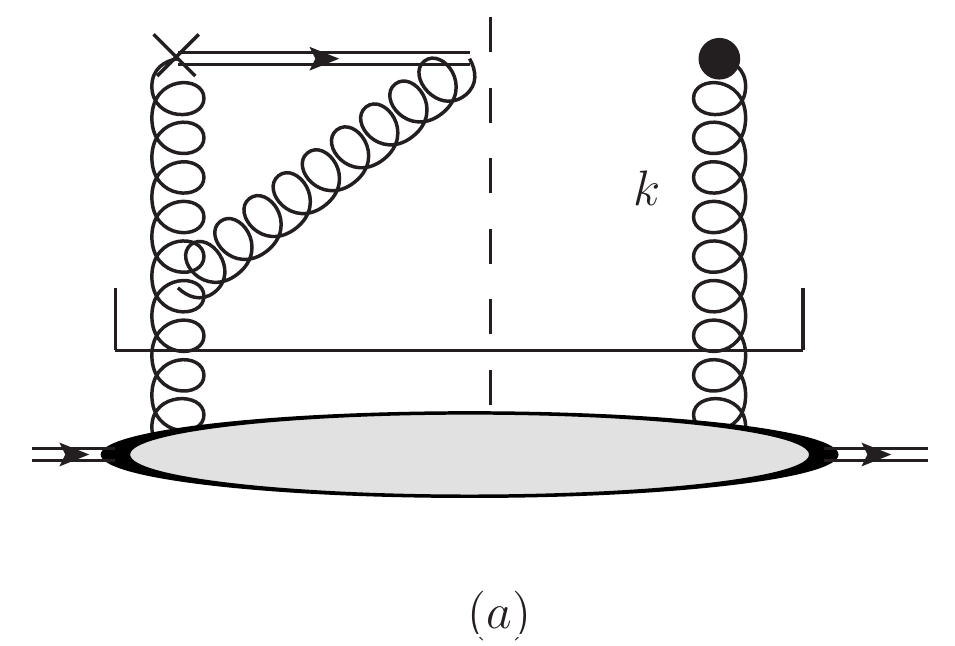}\hspace{2cm}
\includegraphics[scale=0.5]{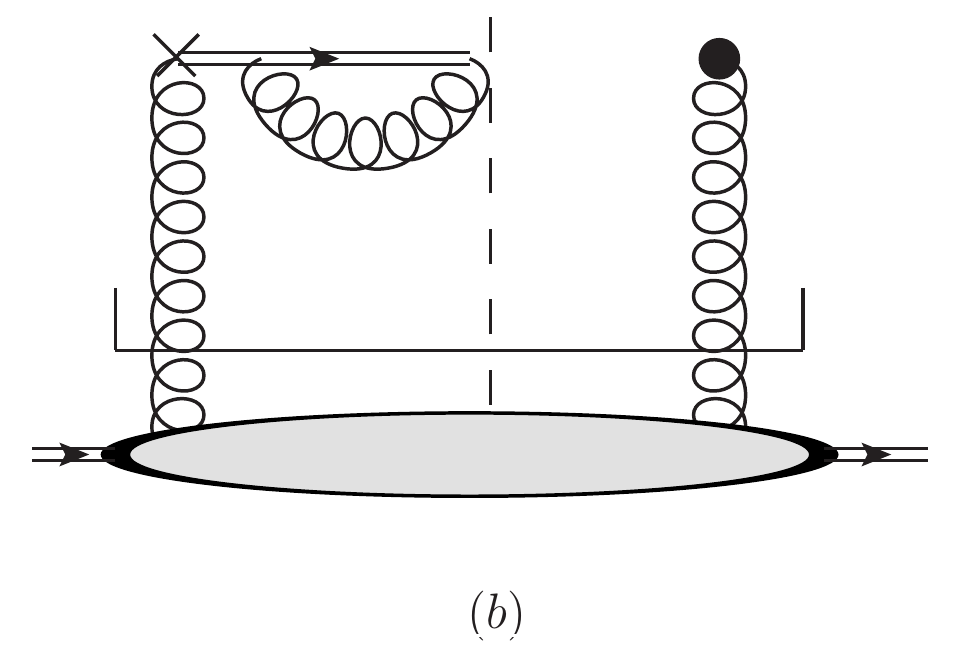}
\caption{Diagrams contributing to virtual correction of gluon TMDPDF, where the cross on left hand side of the cut is the special vertex for gluon field strength tensor. The external gluon takes momentum $k^\mu$, and it is going into the vertex. }
\label{fig:Phi}
\end{figure}
The virtual correction to this function is still obtained by power expansion. The diagrams contributing to virtual correction are given in Fig.\ref{fig:Phi}. Here we take Fig.\ref{fig:Phi}(a) as an example to illustrate our calculation scheme. According to collinear approximation, the leading power contribution of Fig.\ref{fig:Phi}(a) is from the region where the momentum of the parton going through the hooked line is collinear to $P_A$, i.e., $k^\mu=(k^+,k^-,k_\perp)\sim Q(1,\la^2,\la)$. Therefore,
\begin{align}
\Phi^{(1)}_{\al\be}(x,p_\perp)=&\int d^4 k \de^2(k_\perp-p_\perp)\de(k^+-p^+)M_{\al\be}^{\rho\tau,cd}(k^+,k^-,k_\perp)
\int\frac{d^4\xi}{(2\pi)^4}e^{-ik\cdot\xi}\langle P_A|G_{\perp d}^\tau(\xi^+,\xi^-,\xi_\perp)G_{\perp c}^\rho(0)|P_A\rangle\no
\simeq & \int d^2 k_\perp \de^2(k_\perp-p_\perp) M_{\al\be}^{\rho\tau,cd}(p^+,0,k_\perp)
\int\frac{d\xi^- d^2\xi_\perp}{(2\pi)^3}e^{-ip^+\xi^--ik_\perp\cdot \xi_\perp}\langle P_A|G_{\perp b}^\tau(0,\xi^-,\xi_\perp)G_{\perp a}^\rho(0)|P_A\rangle,
\end{align}
where
\begin{align}
M_{\al\be}^{\rho\tau,cd}(k^+,k^-,k_\perp)=&-ig_s^2 \left(\frac{p^+}{2M_p}\right)^{-1}C_A \de_{cd}
\int\frac{d^n l}{(2\pi)^n}\frac{[k^+ g_\perp^{\al\mu}-n^\mu(k+l)_\perp^\al ]v^\nu \Gamma_{\rho\nu\mu}(k,l,-k-l)}{(v\cdot l+i\ep)(l^2+i\ep)[(k+l)^2+i\ep]},\no
\Gamma_{\rho\nu\mu}(k,l,-k-l)=& g_{\rho\nu}(k-l)_\mu+g_{\nu\mu}(2l+k)_\rho+g_{\mu\rho}(-2k-l)_\nu.
\end{align}
Now in $M(k)$ the loop integral is divergent when $k_\perp$ goes to zero if $n=4$. But since the divergence is logarithms-like, it can by regularized in dimensional scheme. Then, $M(k)$ is well-defined at $k_\perp=0$, and it can be expanded as
\begin{align}
M(p^+,0, k_\perp)=& M(p^+,0,0)+k_\perp^\rho \frac{\partial }{\partial k_\perp^\rho}M(p^+,0,0)+\cdots.
\end{align}
Note that the hard scale in $M(p^+,0, k_\perp)$ is $\zeta^2=(2v\cdot p)^2/v^2$, so, high twist contribution is suppressed by $k_\perp/\zeta$.
Since the delta function $\de^2(k_\perp-p_\perp)$ is already of leading power or leading twist, only the first term in the expansion should be preserved. Thus,
\begin{align}
\Phi^{(1)}_{\al\be}(x,p_\perp)\simeq & M(p^+,0,0)_{\al\be}^{\rho\tau,ab}\int d^2 k_\perp \de^2(k_\perp-p_\perp)
\int\frac{d\xi^- d^2\xi_\perp}{(2\pi)^3}e^{-ip^+\xi^--ip_\perp\cdot \xi_\perp}\langle P_A|G_{\perp b}^\tau(0,\xi^-,\xi_\perp)G_{\perp a}^\rho(0)|P_A\rangle.
\end{align}
Now by changing gluon field to field strength tensor the integral of above equation is just gluon TMDPDF itself.
From the derivation one can see that dimensional scheme is crucial for our power expansion. Note that the nonlinear term in gluon field strength tensor also contributes. Its effect is reflected in the special vertex\cite{Collins:1981uw} for the coupling of gluon and gauge link, as shown in Fig.\ref{fig:Phi-special}.
The total correction of Fig.\ref{fig:Phi}(a) is
\begin{align}
\Phi^{(1)}_{\al\be}(x,p_\perp)=&2\times\frac{\al_s C_A}{4\pi}\frac{(4\pi\mu^2)^{\ep/2}}{\Gamma(2-\ep/2)}
\left[\frac{4}{\ep}B(2-\frac{\ep}{2},1+\frac{\ep}{2})
B(-\frac{\ep}{2},1+\frac{\ep}{2})(\zeta^2)^{-\ep/2}+(\frac{2}{\ep_{UV}}-\frac{2}{\ep_{IR}})
\right] \Phi^{(0)}_{\al\be}(x,p_\perp),
\label{eq:Phi-a}
\end{align}
where the factor 2 represents contribution from conjugated diagrams. Note that in this expression and formulas following $\ep$ is $\ep_{IR}$ implicitly, unless there is a special illustration.

Besides, wave function renormalization for the gauge link is given by Fig.\ref{fig:Phi}(b) and its conjugate, the result is
\begin{align}
\Phi^{(1)}_{\al\be}(x,p_\perp)|_{b}=&2\times (Z_v^{1/2}-1)\Phi^{(0)}_{\al\be}(x,p_\perp),\no
Z_v=&1+\frac{\al_s}{\pi}C_A(\frac{1}{\ep_{UV}}-\frac{1}{\ep_{IR}}).
\label{eq:Phi-b}
\end{align}
The sum of eq.(\ref{eq:Phi-a}) and eq.(\ref{eq:Phi-b}) is the total virtual correction to gluon TMDPDF, that is,
\begin{align}
\Phi^{(1)}_{\al\be}(x,p_\perp)=&2W_\Phi \Phi^{(0)}_{\al\be}(x,p_\perp),\no
W_\Phi=&
(Z_v^{1/2}-1)+\frac{\al_s C_A}{4\pi}\frac{(4\pi\mu^2)^{\ep/2}}{\Gamma(2-\ep/2)}
\left[\frac{4}{\ep}B(2-\frac{\ep}{2},1+\frac{\ep}{2})
B(-\frac{\ep}{2},1+\frac{\ep}{2})(\zeta^2)^{-\ep/2}+(\frac{2}{\ep_{UV}}-\frac{2}{\ep_{IR}})
\right].
\label{eq:WPhi}
\end{align}
Since the derivation does not depend on the polarization of initial hadron and the parton or gluon, the result indicates the virtual correction is the same to the two gluon TMDPDFs we consider here. Note that we will not calculate the self-energy correction to initial gluon in our following calculation for one-loop hard part, because this self-energy correction can be subtracted totally. For this reason, we do not calculate this self-energy correction to $\Phi_{\al\be}$ here.

\begin{figure}
\includegraphics[scale=0.5]{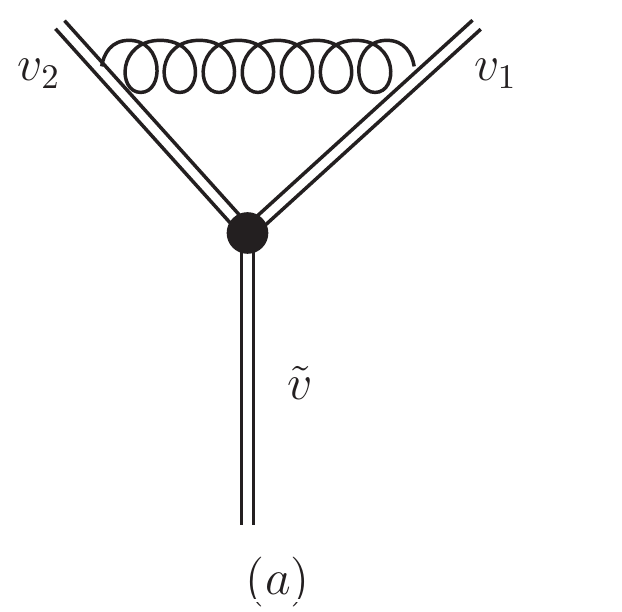}
\includegraphics[scale=0.5]{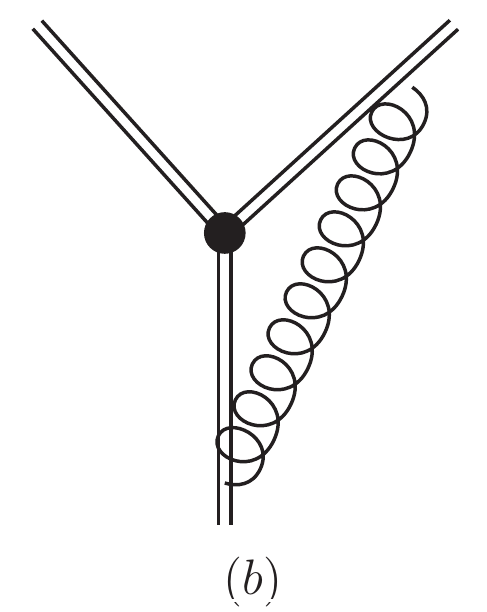}
\includegraphics[scale=0.5]{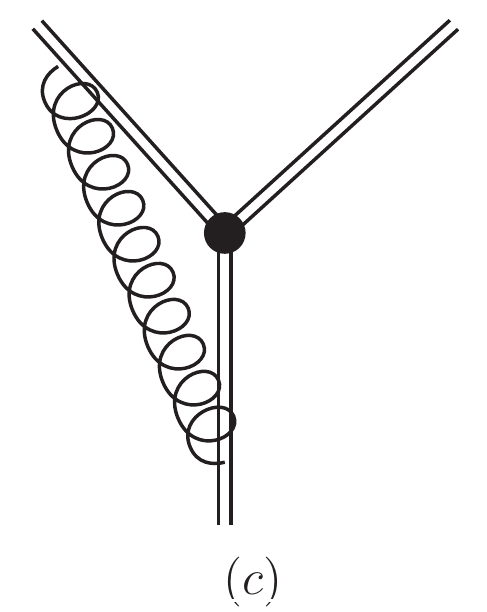}
\includegraphics[scale=0.5]{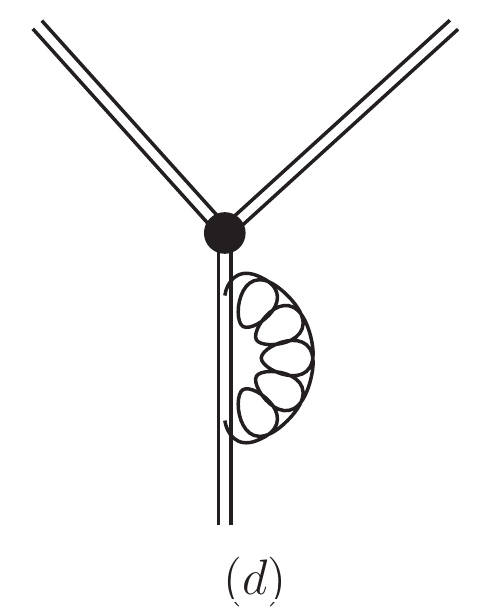}
\includegraphics[scale=0.5]{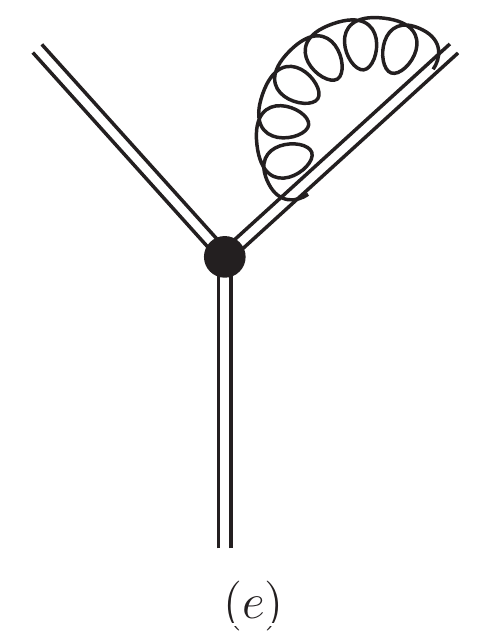}
\includegraphics[scale=0.5]{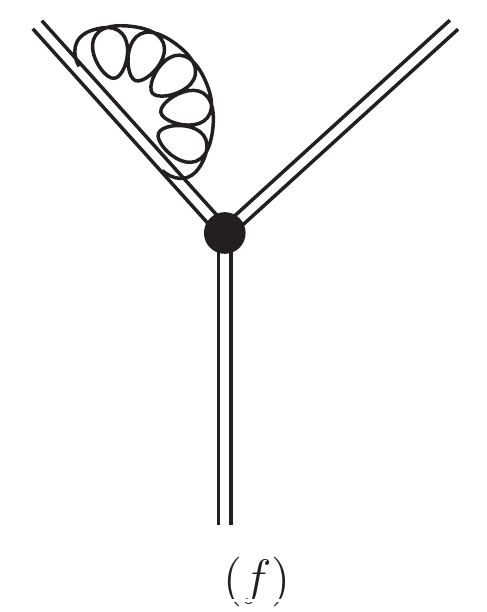}
\caption{Virtual correction to heavy quark soft factor, where we just show the left part of the total diagram, since the right part is trivial. The black dot in the diagram represents color matrix. }
\label{fig:SQ}
\end{figure}

There is no power expansion in higher order correction to soft factors, so, their virtual corrections are easy to calculate. The virtual correction to $S^Q(l_\perp)$ is from Fig.\ref{fig:SQ}.
The result is
\begin{align}
S_Q^{(1)}(l_\perp)=&2W_{SQ} S_Q^{(0)}(l_\perp) ,\no
W_{SQ}=& (\sqrt{Z_{v_1}Z_{v_2}Z_{\tilde{v}}}-1)-\frac{\al_s}{4\pi}\left(\frac{1}{\ep_{UV}}-\frac{1}{\ep_{IR}}\right)
\left[ C_A \left(\ln\frac{(2v_1\cdot\tilde{v})^2}{v_1^2 \tilde{v}^2}+\ln\frac{(2v_2\cdot\tilde{v})^2}{v_2^2 \tilde{v}^2}\right)\right.\no
&\left.+(C_A-2C_F)\frac{1+\la}{\la^{1/2}}\ln\frac{1-\la^{1/2}}{1+\la^{1/2}}
\right],\
\la=\frac{1-\sqrt{v_1^2 v_2^2/(v_1\cdot v_2)^2}}{1+\sqrt{v_1^2 v_2^2/(v_1\cdot v_2)^2}},
\label{eq:WSQ}
\end{align}
$v_1$ and $v_2$ are the four-velocities of heavy quark and anti-quark, respectively. If one takes $v_1^2=v_2^2=1$, $\la$ becomes the usual phase space factor for final heavy quark pair, i.e., $\la=1-4m^2/(k_1+k_2)^2=\rho_{12}^2$.

\begin{figure}
\includegraphics[scale=0.5]{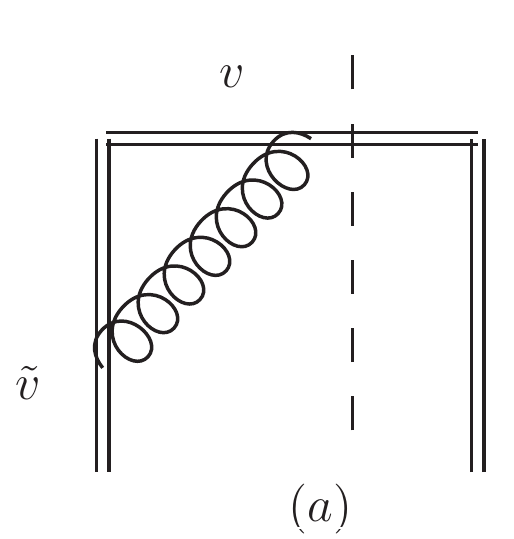}\hspace{1cm}
\includegraphics[scale=0.5]{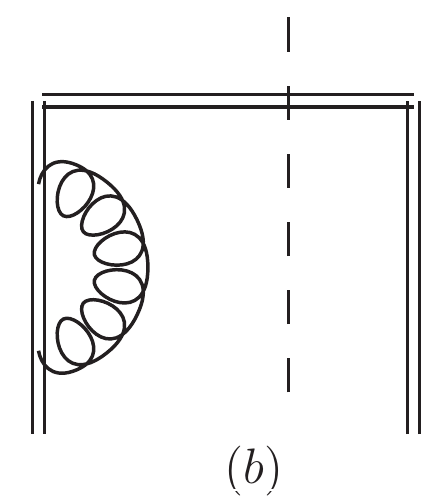}\hspace{1cm}
\includegraphics[scale=0.5]{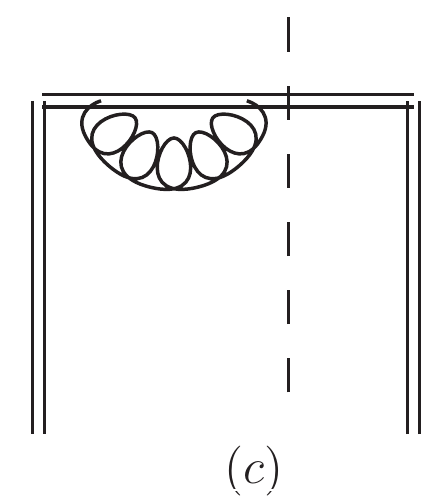}
\caption{Virtual correction to soft factor.}
\label{fig:soft}
\end{figure}
The virtual correction to $S(l_\perp)$ is from Fig.\ref{fig:soft}, and the result is
\begin{align}
\bar{S}^{(1)}(l_\perp)=&2W_S \bar{S}^{(0)}(l_\perp),\no
W_S=&-\left[(\sqrt{Z_v Z_{\tilde{v}}}-1)
-\frac{\al_s}{2\pi}C_A\left(\frac{1}{\ep_{UV}}
-\frac{1}{\ep_{IR}}\right)\ln\frac{(2v\cdot\tilde{v})^2}{v^2\tilde{v}^2}\right].
\label{eq:WS}
\end{align}
Note that in the derivation of these results we have ignored the corrections vanishing under the limit $v^2,\tilde{v}^2\to 0$.

\subsection{Structure of one-loop virtual correction to hard part}
\begin{figure}
\begin{flushleft}
\includegraphics[scale=0.5]{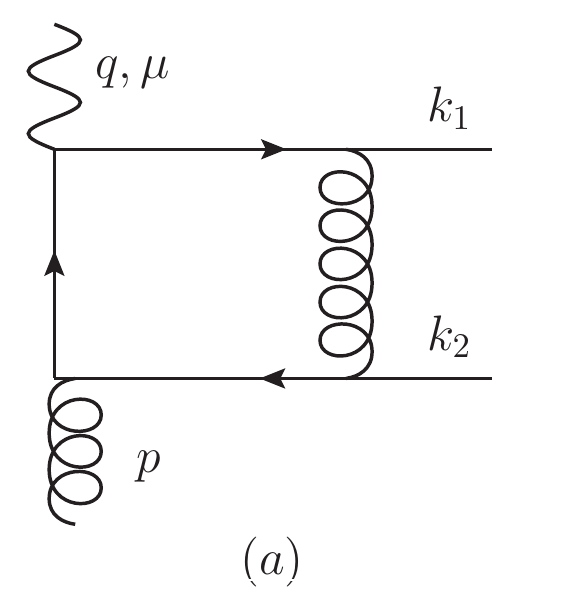}
\includegraphics[scale=0.5]{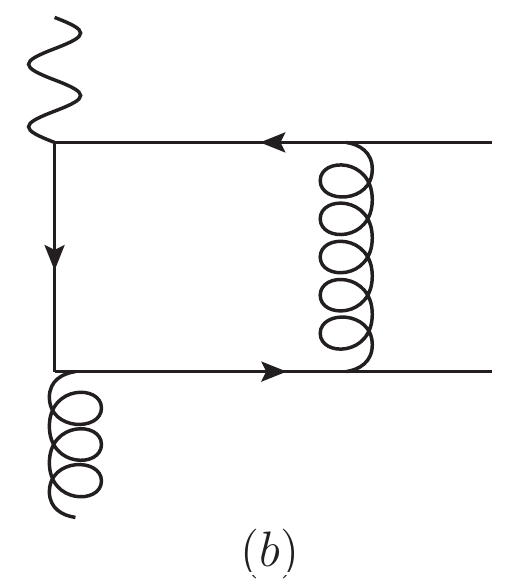}
\includegraphics[scale=0.5]{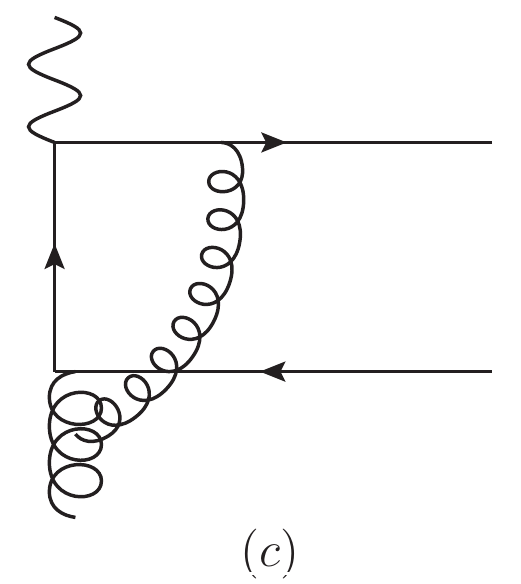}
\includegraphics[scale=0.5]{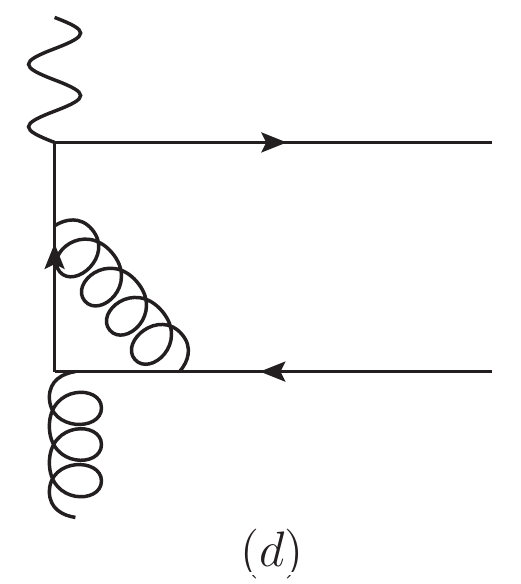}
\includegraphics[scale=0.5]{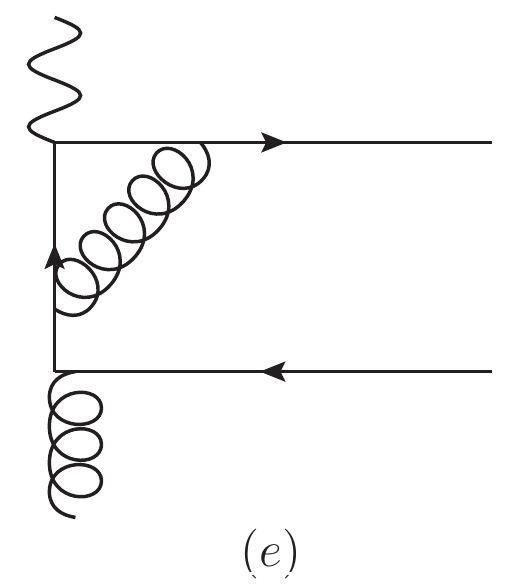}
\includegraphics[scale=0.5]{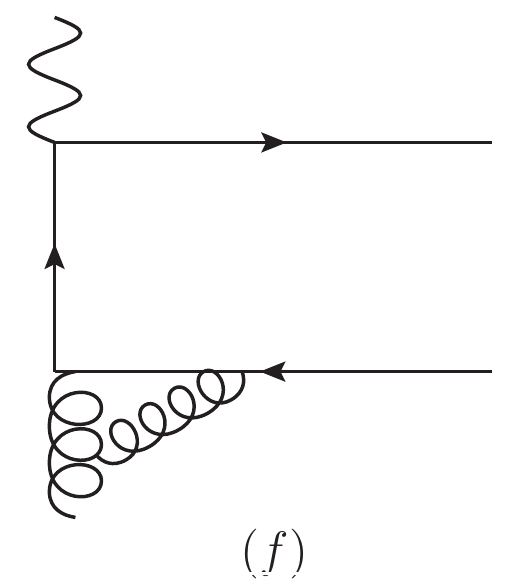}
\includegraphics[scale=0.5]{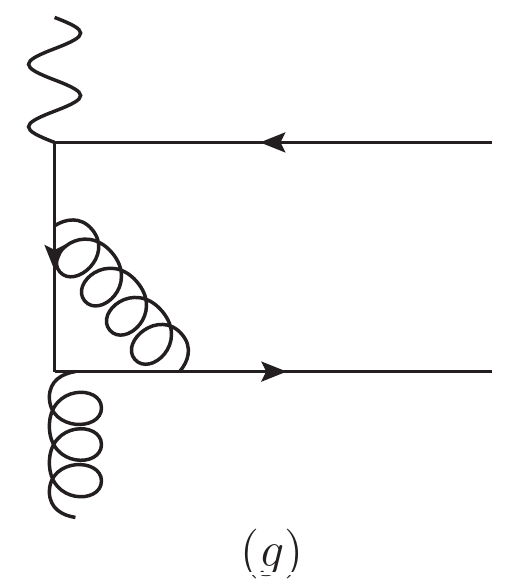}
\includegraphics[scale=0.5]{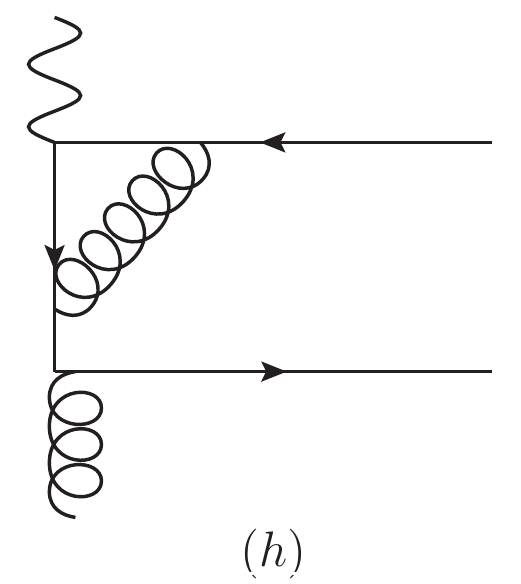}
\includegraphics[scale=0.5]{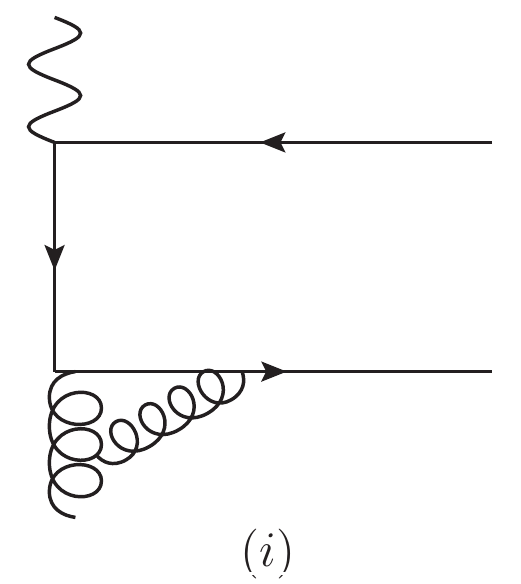}
\includegraphics[scale=0.5]{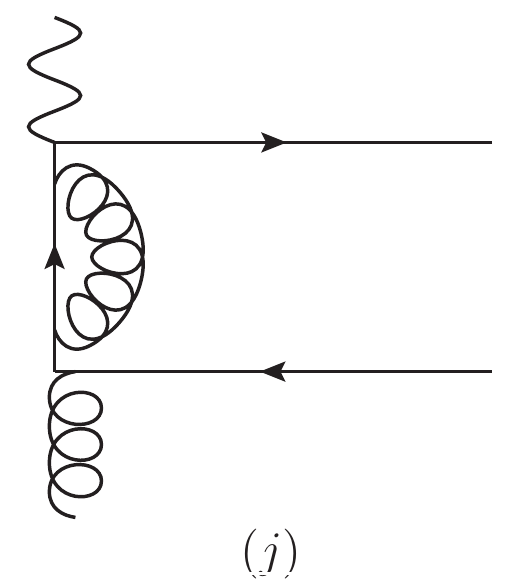}
\includegraphics[scale=0.5]{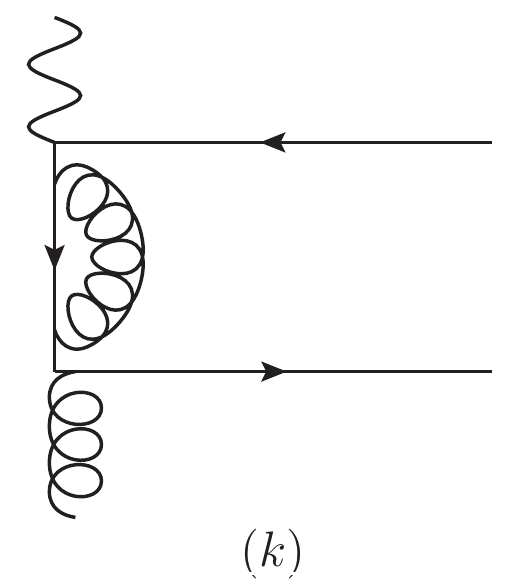}
\end{flushleft}
\caption{One-loop virtual corrections to the amplitude in hard scattering part, where the self-energy corrections to external heavy quark and anti-quark is not shown, since they are trivial to be taken into account.}
\label{fig:oneloop}
\end{figure}

Next we consider the virtual correction to hadronic tensor. The central bubble in Fig.\ref{fig:QQ-prod} on the left hand side of the cut at one-loop level is given by diagrams in Fig.\ref{fig:oneloop}, where self-energy corrections to external fermion lines are not shown, since they are trivial to be taken into account. The contribution of conjugated diagrams are not shown but taken into account in the calculation. Denoting the
amplitude for the subprocess $q(p)+\ga^*(q)\rightarrow Q(k_1)+\bar{Q}(k_2)$ as $\mathcal{M}^\mu_\al$, the one-loop hard part is
\begin{align}
H^{\mu\nu(1)}_{\al\be}=
\mathcal{M}^{\mu(1)}_\al\left( \mathcal{M}^{\nu(0)}_\be \right)^*
+\mathcal{M}^{\mu(0)}_\al\left( \mathcal{M}^{\nu(1)}_\be \right)^*.
\label{eq:oneloop-hard-part}
\end{align}
As an example, the tree level amplitude in Fig.\ref{fig:tree} is
\begin{align}
\mathcal{M}^{\mu(0)}_\al=
-gT^a\left[\frac{1}{2p\cdot k_2}(\bar{u}(k_1)\ga^\mu\s{p}\ga_\al v(k_2)-2k_{2\al} \bar{u}(k_1)\ga^\mu v(k_2))-(k_1\leftrightarrow k_2)\right],
\end{align}
where $\al$ is transverse Lorentz index for initial gluon and $\mu$ is that for photon. By using this amplitude the result in eq.(\ref{eq:hard-coeff-tree})
can be obtained. From eq.(\ref{eq:oneloop-hard-part}) it is clear
\begin{align}
H^{\mu\nu}_{\al\be}=\left(H^{\nu\mu}_{\be\al}\right)^*.
\end{align}
So, if the hard part is symmetric in $(\mu,\al)$ and $(\nu,\be)$, it must be real. Since the transverse momenta $p_\perp$ and $q_\perp$ are ignored in one-loop hard part, the decomposition for tree level hard part in eq.(\ref{eq:hard-coeff}) can also be applied to one-loop hard part. From eq.(\ref{eq:hard-coeff}), the one-loop hard part has such a symmetry automatically. So, all projected hard
coefficients $H^{ij}$ are real. Since tree level amplitude $\mathcal{M}^{(0)}$ is real, we just need the real part of one-loop amplitude. This simplifies the calculation, since absorptive part needs some caution for the $i\ep$ prescription in propagators in the loop.

Next we discuss the IR property of one-loop amplitude. The complete amplitude is very complicated, but the IR divergent part can be shown to be simple and proportional to tree level amplitude. Here IR divergences include soft and collinear ones. These divergences can appear only in Fig.\ref{fig:oneloop}(a,b,c,f,i). Our observation is the loop integrals with IR divergence can be expressed through three basic scalar loop integrals. Fig.\ref{fig:oneloop}(a) contains only soft divergence, which is caused by the soft gluon exchange between the two quarks. Such soft divergence is contained in following integral,
\begin{align}
I_{\tx{Box}-a}^{(0)}(k_1,k_2)=\int_l
\frac{1}{[l^2+i\ep][(l-k_2)^2-m^2+i\ep][(l+p-k_2)^2-m^2+i\ep][(l+k_1)^2-m^2+i\ep]},
\hspace{0.5cm} \int_l=\int\frac{d^nl}{(2\pi)^n}.
\end{align}
The soft divergence of such box integral can be expressed through scalar triangle integrals\cite{Rodrigo:1997gv}. In soft region with $l^\mu=(l^+,l^-,l_\perp^\mu)\sim Q(\la,\la,\la)$, $(l+p-k_2)^2-m^2$ is offshell. Thus setting $l=0$ in this propagator does not affect the IR divergence. This off-shell propagator can be decomposed as
\begin{align}
\frac{1}{(l+p-k_2)^2-m^2}=\frac{1}{(p-k_2)^2-m^2}+\frac{l^2-2l\cdot k_2+2l\cdot p}{2p\cdot k_2[(l+p-k_2)^2-m^2]}.
\end{align}
The second term gives IR finite contribution. Then,
\begin{align}
I_{\tx{Box}-a}^{(0)}(k_1,k_2)=&\frac{1}{-2p\cdot k_2}\tx{CTri}1(s_1)+\tx{DBox}1(t_1,u_1),\no
\tx{CTri}1(s_1)=& \int_l\frac{1}{(l^2+i\ep)[(l-k_2)^2-m^2+i\ep][(l+k_1)^2-m^2+i\ep]}.
\label{eq:DBox1}
\end{align}
All soft divergence of the box integral now is contained in the triangle integral $\tx{CTri}1(s_1)$. Note that $\tx{CTri}1(s_1)$ is symmetric in $k_1$, $k_2$ or $t_1,u_1$. By exchanging $k_1$ and $k_2$, the soft divergence of Fig.\ref{fig:oneloop}(b) is obtained. After some simplifications the sum of the divergent parts of Fig.\ref{fig:oneloop}(a,b) is
\begin{align}
\left.\mathcal{M}^\mu_\al\right|_{a+b}^{IR}\doteq & -ig^2(C_A-2C_F)(2k_1\cdot k_2)\tx{CTri}1(s_1)\left\{
-gT^a\frac{1}{2p\cdot k_2}(\bar{u}(k_1)\ga^\mu\s{p}\ga_\al v(k_2)-2k_{2\al} \bar{u}(k_1)\ga^\mu v(k_2))-(k_1\leftrightarrow k_2)\right\}.
\end{align}
The quantity in $\{\cdots\}$ is rightly the tree level amplitude. Hence,
\begin{align}
\left.\mathcal{M}^\mu_\al\right|_{a+b}^{IR}\doteq & -ig^2(C_A-2C_F)(2k_1\cdot k_2)\tx{CTri}1(s_1)\mathcal{M}^{\mu(0)}_\al.
\label{eq:box-ab}
\end{align}

For Fig.\ref{fig:oneloop}(c), both soft and collinear divergences are contained in the loop integral, and the overlap of these two divergences makes the extraction of IR part nontrivial. The divergent loop integrals for this diagram are
\begin{align}
I_{\tx{Box}-c}^{(0,1,2)}=&\int_l\frac{\{1,(l^+/p^+),(l^+/p^+)^2\}}{D_0 D_1 D_2 D_3},
\end{align}
with
\begin{align}
D_0=l^2,\ D_1=(l+p)^2,\ D_2=(l+p-k_2)^2-m^2,\ D_3=(l+k_1)^2-m^2.
\end{align}
In $I_{\tx{Box}c}^{(0)}$, collinear divergence comes from the region where $l^\mu$ is collinear to $p^\mu$, while soft divergence comes from two regions: one is $l^\mu$ is soft, the other is $(p-l)^\mu$ is soft. In $I_{\tx{Box}c}^{(1,2)}$, collinear divergences come from the same region as $I_{\tx{Box}c}^{(0)}$, but the soft divergence only comes from the region where $(p-l)^\mu$ is soft. Note that the integral with $(l^{+})^3$ in the integrand is absent in the amplitude, although this integral is also IR divergent.

As we will show, the divergent part of Box-c can still be expressed through triangle integrals. First, setting $l=0$ in $D_2$ gives
\begin{align}
I_{\tx{Box}-c}^{(0)}\doteq &-\frac{1}{2p\cdot k_2}\int_l\frac{1}{D_0 D_1 D_3}-\frac{1}{p\cdot k_2}\int_l\frac{l\cdot k_2}{D_0 D_1 D_2 D_3},
\label{eq:boxc-0}
\end{align}
where $\doteq$ means the equality holds up to IR finite corrections. Now $l\cdot k_2=l^+ k_{2}^-+l^- k_2^++l_\perp\cdot k_2$, but $l^-$ and $l_\perp$ are suppressed by $\la$ or $\la^2$ in soft or collinear regions, so, $l^-$ and $l_\perp$ can be dropped. Then
\begin{align}
&\frac{1}{p\cdot k_2}\int_l\frac{l\cdot k_2}{D_0 D_1 D_2 D_3}\doteq\frac{1}{p^+}\int_l\frac{l^+}{D_0 D_1 D_2 D_3}
\doteq \frac{1}{p\cdot q}\int_l\frac{l\cdot q}{D_0 D_1 D_2 D_3}
\doteq \frac{1}{p\cdot q}\int_l\frac{l\cdot (k_1+k_2)}{D_0 D_1 D_2 D_3}\no
= &\frac{1}{p\cdot q}\int_l\frac{D_3-D_0+D_1-D_2-2k_2\cdot p}{D_0 D_1 D_2 D_3},
\end{align}
where we have used $q^\mu=-p^\mu +k_1^\mu +k_2^\mu$. Now $D_{0,1}$ in the numerator above can be dropped, because they are suppressed by $\la$ or $\la^2$ in soft or collinear regions. Then, substituting this result back to eq.(\ref{eq:boxc-0}) one gets
\begin{align}
I_{\tx{Box}-c}^{(0)}= &-\frac{1}{2p\cdot k_2}\int_l\frac{1}{D_0 D_1 D_3}-\frac{1}{2p\cdot k_1}\int_l\frac{1}{D_0 D_1 D_2}+\tx{DBox2}[t_1,u_1].
\label{eq:DBox2}
\end{align}
In this way we express the IR part of Box-c through two triangle integrals.  As expected, this expression is symmetric in $k_1,k_2$. In the above we
have also defined the finite part of this box integral as $\tx{DBox2}[t_1,u_1]$.

Similarly, one can show
\begin{align}
I_{\tx{Box}-c}^{(1)}\doteq &\frac{1}{2p\cdot k_1}\int_l\frac{1}{D_0 D_1 D_2},
\hspace{1cm}
I_{\tx{Box}-c}^{(2)}\doteq -\frac{1}{2p\cdot k_1}\int_l\frac{1}{D_0 D_1 D_2}-\frac{1}{4p\cdot k_1 p\cdot k_2}\int_l\frac{1}{D_0 D_1}.
\end{align}
With the same method, the extraction of IR divergence from the triangle diagrams Fig.\ref{fig:oneloop}(f) becomes easy, and the results are
\begin{align}
I_{\tx{Tri}-f}^{(0)}=&\int_l \frac{1}{l^2(l+p)^2[(l+k_2)^2-m^2]}=\int_l \frac{1}{l^2(l+p)^2[(l+p-k_2)^2-m^2]}=\int_l\frac{1}{D_0 D_1 D_2},\no
I_{\tx{Tri}-f}^{(1)}=&\int_l \frac{(l^+/p^+)}{l^2(l+p)^2[(l+k_2)^2-m^2]}\doteq \frac{1}{2p\cdot k_2}\int_l\frac{1}{D_0 D_1}.
\end{align}
The integrals with $(l^+)^2$ or $(l^+)^3$  do not appear in the amplitude, although they are divergent. Other integrals are IR finite for this triangle diagram. By exchanging $k_1,k_2$ the divergent part of Fig.\ref{fig:oneloop}(i) can be obtained.

With the obtained IR divergence, the amplitude from Fig.\ref{fig:oneloop}(a,b,c,f,i) can be expressed through four basic scalar integrals $J_{01}, J_{012}, J_{013}$ and $\tx{CTri1}(s_1)$. After a lengthy calculation the result is put into a very neat form, which is proportional to tree level amplitude! Besides, the self-energy corrections to external fermion lines also contain IR divergence. Then, the complete IR divergent part of the amplitude is
\begin{align}
\left.\mathcal{M}^{\mu(1)}_\al\right|_{IR}= & W_{IR} \mathcal{M}^{\mu(0)}_\al,\no
W_{IR}=&
(Z_2-1)-ig^2(C_A-2C_F)(2k_1\cdot k_2)\tx{CTri}1(s_1)
-ig^2 C_A(2p\cdot k_1 J_{013}
+2p\cdot k_2 J_{012}+J_{01}),
\label{eq:IR-amp}
\end{align}
with
\begin{align}
J_{012}=\int_l\frac{1}{D_0 D_1 D_2},\ J_{013}=\int_l\frac{1}{D_0 D_1 D_3},\ J_{01}=\int_l\frac{1}{D_0 D_1}.
\end{align}
This is one of our main results. The derivation of this result is independent of the polarizations of external gluon, thus this result indicates the IR divergence, including soft and collinear ones, is spin independent. According to eq.(\ref{eq:oneloop-hard-part}), the IR divergence of hard part is
\begin{align}
\left.H^{\mu\nu(1)}_{\al\be}\right|_{IR}= 2\tx{Re}(W_{IR}) H^{\mu\nu(0)}_{\al\be},
\label{eq:WIR}
\end{align}
where the factor $2$ indicates the contribution from conjugated diagrams.

Besides IR divergence, the hard part also contains UV divergence, which is subtracted by the counter terms of lagrangian. After the subtraction of UV divergence, the hard part
gives a $\mu$ dependence, with $\mu$ the renormalization scale. We will separate the $\mu$ dependence in the following.

Notice that the electro-magnetic current $j^\mu$ in hadronic tensor is conserved, so, the UV divergence from vertex correction is cancelled by self-energy correction to fermions. This means the UV divergences of Fig.\ref{fig:oneloop}(e,h) are cancelled by Fig.\ref{fig:oneloop}(j,k). So, the sum of these four diagrams does not generate $\mu$ dependence. Apart from the $\mu$ dependence in wave function renormalization constant for fermions, i.e., $Z_2$, the remaining $\mu$ dependence can only come from Fig.\ref{fig:oneloop}(d,f,g,i). The UV divergence of these diagrams reads
\begin{align}
\left.\mathcal{M}^\mu_\al\right|_{d+g}\doteq &-\frac{C_A-2C_F}{2}
\frac{\al_s(4\pi\mu^2/Q^2)^{\ep/2}}{4\pi}\Gamma(\frac{\ep_{UV}}{2})
\mathcal{M}^{\mu(0)}_\al,\no
\left.\mathcal{M}^\mu_\al\right|_{f+i}\doteq &\frac{3C_A}{2}
\frac{\al_s(4\pi\mu^2/Q^2)^{\ep/2}}{4\pi}\Gamma(\frac{\ep_{UV}}{2})
\mathcal{M}^{\mu(0)}_\al.
\end{align}
With the $\mu$ dependence from $Z_2$, the total $\mu$-dependence of the amplitude is
\begin{align}
\mathcal{M}^{\mu(1)}_\al=&\left[-\frac{C_A-2C_F}{2}+\frac{3C_A}{2}-C_F\right]
\frac{\al_s}{4\pi}\ln\frac{\mu^2}{Q^2}
\mathcal{M}^{\mu(0)}_\al+\cdots,
\end{align}
where $\cdots$ represents $\mu$ independent part.

With renormalization scale separated explicitly, the hard part can be written as
\begin{align}
H^{\mu\nu(1)}_{\al\be}=H^{\mu\nu}_{1\al\be}+H^{\mu\nu}_{2\al\be}+H^{\mu\nu}_{3\al\be},
\end{align}
with
\begin{align}
H^{\mu\nu}_{1\al\be}=&\left.2\tx{Re}(W_{IR}) H^{\mu\nu(0)}_{\al\be}\right|_{\mu^2=Q^2},\no
H^{\mu\nu}_{2\al\be}=&\left(H^{\mu\nu(1)}_{\al\be}-
2\tx{Re}(W_{IR}) H^{\mu\nu(0)}_{\al\be}\right)_{\mu^2=Q^2},\no
H^{\mu\nu}_{3\al\be}=&
2\frac{\al_s}{4\pi}C_A\ln\frac{\mu^2}{Q^2}H^{\mu\nu(0)}_{\al\be},
\label{eq:H-123}
\end{align}
where we have chosen the reference scale of $\mu^2$ as $Q^2$ so that $H_3$ is definite. These three parts make the structure of one-loop virtual correction to hadronic tensor clear.

At last, it should be pointed out that only in pole mass scheme, the $\mu$ dependence of Fig.\ref{fig:oneloop}(j,k) is proportional to tree level amplitude. In pole mass scheme the mass counter term in lagrangian is chosen so that renormalized fermion self-energy
$\Sigma_R(\s{\tilde{p}},m)=0$ when $\s{\tilde{p}}=m$. Then the renormalized self-energy is written as
\begin{align}
\Sigma_R(\tilde{\s{p}},m)=&-\frac{\al_s}{4\pi}C_F\ln\frac{\mu^2}{Q^2}(\tilde{\s{p}}-m)+\frac{\al_s}{4\pi}C_F\left(\tilde{\s{p}}B_v(\frac{-\tilde{p}^2}{m^2},\frac{Q^2}{m^2})
+m B_m(\frac{-\tilde{p}^2}{m^2},\frac{Q^2}{m^2})\right),
\end{align}
with
\begin{align}
B_v(z,\tau)&=\int_0^1 dx x\ln(x z+1)-\ln\tau=\frac{(2-z)z-2(1-z^2)\ln(1+z)}{4z^2}-\ln\tau,\no
B_m(z,\tau)&=-2-4\int_0^1 dx\ln(xz+1)+\ln\tau=2-\frac{4(1+z)}{z}\ln(1+z)+\ln\tau.
\label{eq:BvBm}
\end{align}
From the structure of self-energy correction, $\ln\mu$ part is proportional to tree diagrams. As stated before this $\mu$ dependence is cancelled by that of Fig.\ref{fig:oneloop}(e,h). This is an advantage of pole mass scheme.

\subsection{Subtraction and finite hard coefficients}\label{sec:finite-hard-part}
According to our subtraction scheme in eq.(\ref{eq:subtraction-scheme}) and results in eq.(\ref{eq:WPhi},\ref{eq:WSQ},\ref{eq:WS},\ref{eq:WIR}), the finite hard coefficient is given by
\begin{align}
\left.H^{\mu\nu(1)}_{\al\be}\right|_{finite}=& H^{\mu\nu(1)}_{\al\be}
-2Re[W_\Phi+W_{SQ}+W_{S}]H^{\mu\nu(0)}_{\al\be}\no
=& 2Re\left[\left(W_{IR}+C_F\frac{\al_s}{4\pi}\ln\frac{\mu^2}{Q^2}
-C_A\frac{\al_s}{4\pi}\ln\frac{\mu^2}{Q^2}\right)
-W_\Phi-W_{SQ}-W_{S}\right]H^{\mu\nu(0)}_{\al\be}
+H^{\mu\nu}_{3\al\be}+H^{\mu\nu}_{2\al\be},
\label{eq:Hfinite}
\end{align}
where we have used the fact
\begin{align}
W_{IR}|_{\mu^2=Q^2}=W_{IR}+C_F\frac{\al_s}{4\pi}\ln\frac{\mu^2}{Q^2}
-C_A\frac{\al_s}{4\pi}\ln\frac{\mu^2}{Q^2},
\end{align}
and the $\ln\mu^2$ is from $Z_2$ and $J_{01}$.

The explicit expression of IR divergent scalar loop integrals are calculated by standard Feynman parametrization. One can also find general expressions in e.g. \cite{Rodrigo:1997gv},\cite{Ellis:2007qk} and then continue the result into DIS region. Generally, the result in DIS region is very simple. For completeness we list the result of these integrals here as
\begin{align}
\tx{CTri}1(s_1)= &-\frac{i}{16\pi^2}R_\ep
\frac{1}{s_{12}\rho_{12}}\left\{
-\frac{2}{\ep}\ln c_{12}
+(1-\ln\frac{Q^2}{s_{12}})\ln c_{12}
+\frac{1}{2}\ln^2c_{12}\right.\no
&\left.+2\ln\rho_{12}\ln c_{12}
+\frac{2\pi^2}{3}+2Li_2(c_{12})\right\},\hspace{0.5cm} R_\ep=\frac{(4\pi\mu^2/Q^2)^{\ep/2}}{\Gamma(2-\ep/2)},
\end{align}
and
\begin{align}
J_{013}=& \frac{i}{16\pi^2}R_\ep\frac{1}{Q^2 y_{13}}\frac{2}{\ep^2}\left[ 1-\frac{\ep}{2}
+\frac{\ep}{2}\ln\frac{r_b}{y_{13}^2}+\frac{\ep^2}{2}\left(-\frac{1}{2}\ln\frac{r_b}{y_{13}^2}
+\frac{1}{4}\ln^2\frac{r_b}{y_{13}^2}+\frac{\pi^2}{12}+Li_2(\frac{r_b+y_{13}}{r_b})\right)\right],\no
J_{012}=& \left. J_{013}\right|_{y_{13}\to y_{23}},\no
J_{01}=& \frac{i}{16\pi^2}\left(\frac{2}{\ep_{UV}}-\frac{2}{\ep_{IR}}\right),
\end{align}
where
\begin{align}
s_{12}=(k_1+k_2)^2,\ \rho_{12}=\sqrt{1-\frac{4m^2}{s_{12}}},
\ c_{12}=\ln\frac{1-\rho_{12}}{1+\rho_{12}},
\end{align}
and
\begin{align}
y_{13}=\frac{2p\cdot k_1}{q^2},\ y_{23}=\frac{2p\cdot k_2}{q^2},\
y_{12}=-\frac{2k_1\cdot k_2}{q^2},\ r_b=-\frac{m^2}{q^2}.
\end{align}
In DIS region, $y_{13},y_{23}<0$ and $y_{12},r_b>0$. In these integrals we have ignored all absorptive parts.

With these integrals, the explicit $W_{IR}$ can be obtained as
\begin{align}
W_{IR}=& (Z_2-1)-(C_A-2C_F)\frac{\al_s}{4\pi}R_\ep \frac{1+\rho_{12}^2}{2\rho_{12}}
\left\{
-\frac{2}{\ep}\ln c_{12}
+(1-\ln\frac{Q^2}{s_{12}})\ln c_{12}
+\frac{1}{2}\ln^2c_{12}\right.\no
&\left.+2\ln\rho_{12}\ln c_{12}
+\frac{2\pi^2}{3}+2Li_2(c_{12})\right\}\no
&-C_A \frac{\al_s}{4\pi}R_\ep \left\{ \frac{4}{\ep^2}-\frac{2}{\ep}+\frac{2}{\ep}\ln\frac{r_b}{y_{13}y_{23}}
-\ln\frac{r_b}{y_{13}y_{23}}\right.\no
&\left.+\frac{\pi^2}{6}
+\frac{1}{4}\ln^2\frac{r_b}{y_{13}^2}+\frac{1}{4}\ln^2\frac{r_b}{y_{23}^2}
+Li_2(\frac{r_b+y_{13}}{r_b})+Li_2(\frac{r_b+y_{23}}{r_b})\right\}\no
&+C_A\frac{\al_s}{4\pi}\left(\frac{2}{\ep_{UV}}-\frac{2}{\ep_{IR}}\right).
\end{align}
The expression of $W_\Phi$ has been given by eq.(\ref{eq:WPhi}), then we have
\begin{align}
W_{IR}-W_\Phi=&(Z_2-Z_v^{1/2})-(C_A-2C_F)\frac{\al_s}{4\pi}R_\ep \frac{1+\rho_{12}^2}{2\rho_{12}}
\left\{
-\frac{2}{\ep}\ln c_{12}
+(1-\ln\frac{Q^2}{s_{12}})\ln c_{12}
+\frac{1}{2}\ln^2c_{12}\right.\no
&\left.+2\ln\rho_{12}\ln c_{12}
+\frac{2\pi^2}{3}+2Li_2(c_{12})\right\}\no
&-C_A\frac{\al_s}{4\pi}R_\ep\left\{
(\frac{2}{\ep}-1)\ln\frac{(v\cdot\tilde{v})^2}{v^2 v_1\cdot\tilde{v}v_2\cdot \tilde{v}}-\frac{\pi^2}{6}+\frac{1}{4}\ln^2\frac{r_b}{y_{13}^2}
+\frac{1}{4}\ln^2\frac{r_b}{y_{23}^2}-\frac{1}{2}\ln^2\frac{Q^2}{\zeta^2}\right.\no
&\left.+Li_2(\frac{r_b+y_{13}}{r_b})+Li_2(\frac{r_b+y_{23}}{r_b})
\right\}.
\end{align}
It can be seen the IR divergence from hard part cannot be subtracted by the correction to gluon TMDPDF alone. The TMD formula without soft factors must be wrong. At one-loop level, the two soft factors have been calculated in eq.(\ref{eq:WSQ}) and eq.(\ref{eq:WS}). We have
\begin{align}
W_{SQ}+W_S=& (\sqrt{Z_{v_1}Z_{v_2}Z_{\tilde{v}}}-\sqrt{Z_v Z_{\tilde{v}}})\no
&-\frac{\al_s}{4\pi}\left(\frac{1}{\ep_{UV}}-\frac{1}{\ep_{IR}}\right)
\left[2C_A \ln\frac{v^2 v_1\cdot\tilde{v} v_2\cdot\tilde{v}}{(v\cdot\tilde{v})^2}
+(C_A-2C_F)\frac{1+\rho_{12}^2}{\rho_{12}}\ln\frac{1-\rho_{12}}{1+\rho_{12}}\right].
\end{align}
Now it is very clear that the difference between $W_{IR}-W_\Phi$ and $W_{SQ}+W_S$ is IR finite, i.e.,
\begin{align}
&(W_{IR}-W_\Phi)-(W_{SQ}+W_S)\no
=&-\frac{\al_s}{4\pi}C_F(4+3\ln\frac{\mu^2}{m^2})\no
&-(C_A-2C_F)\frac{\al_s}{4\pi}\frac{1+\rho_{12}^2}{2\rho_{12}}
\left\{
-\ln\frac{\mu^2}{Q^2}\ln c_{12}
+(1-\ln\frac{Q^2}{s_{12}})\ln c_{12}
+\frac{1}{2}\ln^2c_{12}\right.\no
&\left.+2\ln\rho_{12}\ln c_{12}
+\frac{2\pi^2}{3}+2Li_2(c_{12})\right\}\no
&-C_A\frac{\al_s}{4\pi}\left\{
(\ln\frac{\mu^2}{Q^2}-1)\ln\frac{(v\cdot\tilde{v})^2}{v^2 v_1\cdot\tilde{v}v_2\cdot \tilde{v}}-\frac{\pi^2}{6}+\frac{1}{4}\ln^2\frac{r_b}{y_{13}^2}
+\frac{1}{4}\ln^2\frac{r_b}{y_{23}^2}-\frac{1}{2}\ln^2\frac{Q^2}{\zeta^2}\right.\no
&\left.+Li_2(\frac{r_b+y_{13}}{r_b})+Li_2(\frac{r_b+y_{23}}{r_b})
\right\},
\label{eq:Wdifference}
\end{align}
In this result, the UV counter terms for gluon TMDPDF and soft factors have been added and the scale $\mu$ is for UV renormalization.
In eq.(\ref{eq:Wdifference}) the first line in the equality is given by wave function renormalization, that is,
\begin{align}
(W_{IR}-W_\Phi)-(W_{SQ}+W_S)\supset Z_2-\sqrt{Z_{v_1}Z_{v_2}}.
\end{align}
After UV divergences are removed the wave function renormalization constants are
\begin{align}
Z_2=&1-\frac{\al_s}{4\pi}C_F\ln\frac{\mu^2}{Q^2}-\frac{\al_s}{2\pi}C_F\left(\frac{2}{\ep_{IR}}-\ga_E+2+\ln 4\pi+\frac{3}{2}\ln\frac{Q^2}{m^2}+\ln\frac{\mu_{IR}^2}{Q^2}\right),\no
Z_{v_1}=Z_{v_2}=&1+\frac{\al_s}{2\pi}C_F(\frac{2}{\ep_{UV}}-\frac{2}{\ep_{IR}})
+\tx{c.t.}\no
=& -\frac{\al_s}{2\pi}C_F(\frac{2}{\ep_{IR}}-\ga_E+\ln \frac{4\pi\mu_{IR}^2}{\mu^2}).
\end{align}
So,
\begin{align}
Z_2-\sqrt{Z_{v_1}Z_{v_2}}=-\frac{\al_s}{4\pi}C_F(4+3\ln\frac{\mu^2}{m^2}).
\end{align}

Now, substituting eq.(\ref{eq:Wdifference}) into eq.(\ref{eq:Hfinite}), the finite hard coefficient is
\begin{align}
\left.H^{\mu\nu(1)}_{\rho\tau}\right|_{finite}=& W_f H^{\mu\nu(0)}_{\rho\tau}+H^{\mu\nu}_{2\rho\tau},\no
W_f=&2\left\{
\frac{\al_s}{4\pi}\ln\frac{\mu^2}{Q^2}\left[-C_A\ln\frac{(v\cdot \tilde{v})^2}{v^2 \tilde{v}\cdot v_1 \tilde{v}\cdot v_2}-2C_F
+(C_A-2C_F)\left(\frac{1+\rho_{12}^2}{2\rho_{12}}\ln\frac{1-\rho_{12}}{1+\rho_{12}}
\right)\right]\right.\no
&\left.+\left(W_{IR}-W_{\Phi}-W_{SQ}-W_S\right)_{\mu^2=Q^2}\right\}\no
=&\frac{\al_s}{2\pi}\ln\frac{\mu^2}{Q^2}\left[-C_A\ln\frac{(v\cdot \tilde{v})^2}{v^2 \tilde{v}\cdot v_1 \tilde{v}\cdot v_2}-2C_F
+(C_A-2C_F)\left(\frac{1+\rho_{12}^2}{2\rho_{12}}\ln\frac{1-\rho_{12}}{1+\rho_{12}}
\right)\right]\no
&-(C_A-2C_F)\frac{\al_s}{2\pi}\frac{1+\rho_{12}^2}{2\rho_{12}}
\left[
(1-\ln\frac{Q^2}{s_{12}})\ln c_{12}
+\frac{1}{2}\ln^2c_{12}
+2\ln\rho_{12}\ln c_{12}
+\frac{2\pi^2}{3}+2Li_2(c_{12})\right]\no
&-C_A\frac{\al_s}{2\pi}\left[
-\frac{\pi^2}{6}+\frac{1}{4}\ln^2\frac{r_b}{y_{13}^2}
+\frac{1}{4}\ln^2\frac{r_b}{y_{23}^2}-\frac{1}{2}\ln^2\frac{Q^2}{\zeta^2}
+Li_2(\frac{r_b+y_{13}}{r_b})+Li_2(\frac{r_b+y_{23}}{r_b})
\right]\no
&-\frac{\al_s}{2\pi}C_F(4+3\ln\frac{Q^2}{m^2}).
\label{eq:hard-part-result}
\end{align}
Moreover, our decomposition for hard coefficient in eq.(\ref{eq:hard-coeff}) is effective to all orders of $\al_s$ if only virtual correction is involved. So, $H^{\mu\nu}_{2\al\be}$
can still be projected into 10 scalar hard coefficients $H_2^{ij}$ as we done for $H^{\mu\nu(0)}_{\al\be}$. There will be no other new tensor structure in higher order correction. That is,
\begin{align}
H^{(1)ij}_{finite}=W_f H^{(0)ij}+H_2^{ij},
\label{eq:hard-part-projected}
\end{align}
with the projection in eq.(\ref{eq:hard-coeff}). There is not any new tensor structure appearing in higher order correction, since only virtual correction contributes. Eq.(\ref{eq:hard-part-result},\ref{eq:hard-part-projected}) is one of our main results. From the derivation, $W_{IR}$ and $W_\Phi$ are independent of the polarizations of parton or gluon, so, $W_f$ is the same to all
gluon TMDPDFs, including those defined by polarized hadron as given in \cite{Mulders:2000sh}. $H_2$ depends on the polarizations of gluon, but it is IR finite and $\mu$ independent. The finiteness of $H_2$ is justified explicitly according to eq.(\ref{eq:H-123}). In our result, it is expressed through some basic IR finite scalar loop integrals, including box integrals $\tx{DBox}1,2$, triangle integrals $\tx{CTri}3,4,5$ and various bubble integrals $b_0$, $B_v$ and $B_m$. These integrals are illustrated in Appendix..

In our result, the projected hard coefficients $H_2^{ij}$ according to eq.(\ref{eq:hard-coeff}) are given in a mathematica file
$subted2.m$. These results are very lengthy and cannot be shown here.
In $subted2.m$ these projected hard coefficients are stored as a list
\begin{align}
\{\xi_1 H_2^{11},\xi_2 H_2^{12},\xi_3 H_2^{21},\xi_4 H_2^{22},\xi_5 H_2^{33},\xi_6 H_2^{41},\xi_7 H_2^{42},\xi_8 H_2^{51},\xi_9 H_2^{52},\xi_{10} H_2^{63}\},
\end{align}
with normalization factors
\begin{align}
\xi_1=&\xi_2=\frac{1}{g_s^4}Q^2|k_{1\perp}|^2,\no
\xi_3=&\xi_4=\xi_5=-\frac{1}{g_s^4}2Q|k_{1\perp}|^3,\no
\xi_6=&\xi_7=\xi_8=\xi_9=\xi_{10}=\frac{1}{g_s^4}2|k_{1\perp}|^4.
\end{align}
All of these results are expressed through the invariants $s_1,t_1,u_1$ and $m$ appearing in tree level result eq.(\ref{eq:hard-coeff-tree}).
There are two color factors in the stored result, i.e.,
\begin{align}
f(N_1)=Tr(T^aT^bT^aT^b)=\frac{1-N_c^2}{4N_c},\ f(N_2)=Tr(T^aT^aT^bT^b)=\frac{(N_c^2-1)^2}{4N_c}.
\end{align}

In the final part it is interesting to point out that our factorization formula eq.(\ref{eq:oneloop-formula})
associated with the finite hard part given by eq.(\ref{eq:hard-part-result}) is $\mu$ independent to order
of $\al_s^2$. This can be illustrated in the following. Because real correction is UV finite, $\mu$ dependence is purely generated by virtual corrections. Then, the $\mu$
dependence of $S^Q(l_\perp)$ and $\bar{S}(l_\perp)$ can be obtained from eq.(\ref{eq:WSQ}) and eq.(\ref{eq:WS}), respectively. They are
\begin{align}
\frac{\partial S^Q(l_\perp)}{\partial\ln\mu^2}
=&\frac{\al_s}{2\pi}\left[2C_F+C_A
-C_A\ln\frac{4v_1\cdot\tilde{v}v_2\cdot\tilde{v}}{\tilde{v}^2}
-(C_A-2C_F)\frac{1+\rho_{12}^2}{2\rho_{12}}\ln\frac{1-\rho_{12}}{1+\rho_{12}}
\right]S^Q(l_\perp),\no
\frac{\partial \bar{S}(l_\perp)}{\partial\ln\mu^2}
=&\frac{\al_s}{2\pi}C_A\left[
-2+\ln\frac{4(v\cdot\tilde{v})^2}{v^2\tilde{v}^2}
\right]\bar{S}(l_\perp).
\end{align}
For gluon TMDPDF, additional self-energy correction to gluon with momentum $p$ should be added into virtual correction. Of course, this correction does not affect the finite hard part, because it cancels the same gluon self-energy insertion in hard scattering for hadronic tensor. The wave function renormalization constant in Feynman gauge is
\begin{align}
Z_G=1-\frac{\al_s}{4\pi}\left(\frac{2}{3}N_F-\frac{5}{3}C_A\right)
(\frac{2}{\ep_{UV}}-\ga_E+\ln 4\pi),
\end{align}
in $\overline{MS}$ scheme. Then, from eq.(\ref{eq:WPhi}) we have
\begin{align}
\frac{\partial }{\partial\ln\mu^2}\Phi^{\al\be}(x,p_\perp)
=&-\frac{\al_s}{4\pi}\left[ \frac{2}{3}N_F-\frac{5}{3}C_A-4C_A
\right]\Phi^{\al\be}(x,p_\perp).
\end{align}
Recall that tree level hard coefficients are proportional to $\al_s$, which has a $\mu$ dependence
\begin{align}
\frac{\partial\al_s}{\partial\ln\mu^2}=\beta(\al_s)=
-\frac{\al_s^2}{4\pi}\left(\frac{11}{3}C_A-\frac{2}{3}N_F \right).
\end{align}
Now the hadronic tensor to $\mathcal{O}(\al_s^2)$ can be written as
\begin{align}
W^{\mu\nu}\propto& \al_s\left[1+\frac{\al_s}{2\pi}\ln\frac{\mu^2}{Q^2}
\left(-C_A\ln\frac{(v\cdot\tilde{v})^2}{v^2\tilde{v}\cdot v_1 \tilde{v}\cdot v_2}-2C_F+(C_A-2C_F)\frac{1+\rho_{12}^2}{2\rho_{12}}
\ln\frac{1-\rho_{12}}{1+\rho_{12}}
\right)+\cdots
\right]\no
&\Phi^{\al\be}(x,p_\perp)\otimes S^Q(l_{1\perp})\otimes\bar{S}(l_{2\perp}),
\end{align}
where $\cdots$ part is $\mathcal{O}(\al_s)$ and $\mu$ independent; $\otimes$ means the transverse momentum convolution. From this equation and evolution equations given above, one can check quickly
\begin{align}
\frac{\partial}{\partial\ln\mu^2}W^{\mu\nu}=0+\mathcal{O}(\al_s^3).
\end{align}
So, the factorization formula is $\mu$ independent to order $\al_s^2$. Notice that here we have used the fact that pole mass of heavy quark is $\mu$ independent.

\section{Azimuthal angle dependence}
Now we have checked the corrected TMD formula for hadronic tensor, i.e., eq.(\ref{eq:oneloop-formula}) at one-loop level, and the finite hard coefficients are given explicitly. One may think that the azimuthal angle dependence about $\phi_q$ can be obtained by following the same procedure as tree level formula. But unfortunately, $S^Q$ depends on $v_{1\perp}$, which will change the tree level $\phi_q$ dependence in a nonperturbative way after integration over the transverse momenta in eq.(\ref{eq:oneloop-formula}). Physically, this change of $\phi_q$ dependence is caused by the multiple emissions of soft gluons from final heavy quark pair. To proceed we have to integrate over $\phi_q$, and it is best to do this in coordinate space.

\begin{figure}
\includegraphics[scale=0.5]{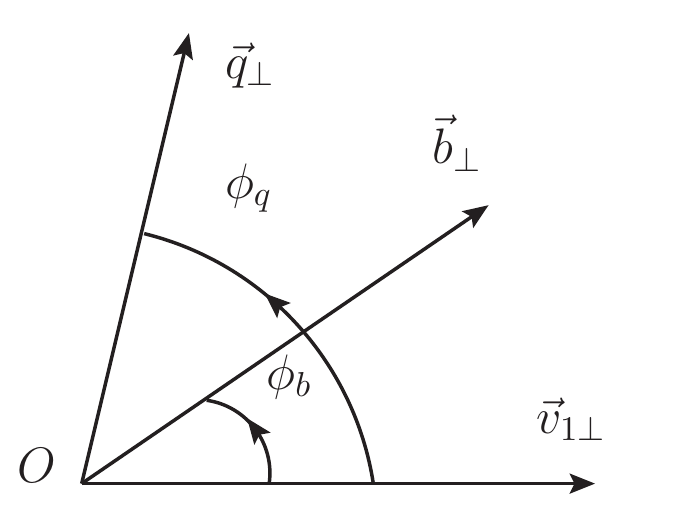}
\caption{Azimuthal angles $\phi_b$ and $\phi_q$ in hadron frame.}
\label{fig:angle-bq}
\end{figure}

Define
\begin{align}
\tilde{\Phi}^{\al\be}(x,b_\perp)=&\int d^2 p_\perp e^{ib_\perp\cdot p_\perp}\Phi^{\al\be}(x,p_\perp)=-g_\perp^{\al\be} G(x,b_\perp^2)
-\frac{2(2b_\perp^\al b_\perp^\be-b_\perp^2 g_\perp^{\al\be})}{M_p^2}\frac{\partial^2 H^\perp(x,b_\perp^2)}{\partial(b_\perp^2)^2},\no
\tilde{S}^Q(b_\perp)=& \int d^2 l_\perp e^{-ib_\perp\cdot l_\perp} S^Q(l_\perp),\no
\tilde{S}(b_\perp)=& \int d^2 l_\perp e^{-ib_\perp\cdot l_\perp} \bar{S}(l_\perp).
\end{align}
Note that $b_\perp^2=b_\perp\cdot b_\perp<0$.
The formula eq.(\ref{eq:oneloop-formula}) becomes
\begin{align}
W^{\mu\nu}
=&\frac{x_B}{x Q^2 M_p(N_c^2-1)}\de(1-z_1-z_2)H^{\mu\nu}_{\al\be}\int \frac{d^2 b_\perp}{(2\pi)^2}e^{iq_\perp\cdot b_\perp}\tilde{\Phi}^{\al\be}(x,b_\perp)\tilde{S}^Q(b_\perp)\tilde{S}(b_\perp^2).
\end{align}
Remember that in hadron frame, i.e., the rest frame of final heavy quark pair, $\vec{v}_{1\perp}$ is along $+X$-axis, then all azimuthal angles are relative to $\vec{v}_{1\perp}$,
as shown in Fig.\ref{fig:angle-bq}.


Since $L^{\mu\nu}$ does not depend on $\phi_q$, it is convenient to integrate over $\phi_q$ in $W^{\mu\nu}$ rather than the cross section. Notice that due to scaling invariance of
$S^Q(b_\perp)$ under the transformation $v_1\to \la_1 v_1$, $v_2\to \la_2 v_2$, $\tilde{S}^Q$ depends on $\phi_b$ through
\begin{align}
\frac{(b_\perp\cdot v_1)^2}{b_\perp^2 v_1^2}=-|v_{1\perp}|^2 \cos^2\phi_b=-|v_{1\perp}|^2 \frac{\cos 2\phi_b+1}{2}.
\end{align}
That is, $\tilde{S}^Q$ is a function of $\cos 2\phi_b$. So, $\tilde{S}^Q$ is unchanged under the transformation $\phi_b\to \phi_b+\pi$. Due to this fact we have
\begin{align}
\int_0^{2\pi} d\phi_b \cos n\phi_b \tilde{S}^Q(b_\perp)=&(-1)^n \int_0^{2\pi} d\phi_b \cos n\phi_b \tilde{S}^Q(b_\perp),\hspace{0.5cm}
\int_0^{2\pi} d\phi_b \sin n\phi_b \tilde{S}^Q(b_\perp)=0.
\end{align}
This property of $\tilde{S}^Q$ is valuable in our following analysis. We choose to study three quantities with $\phi_q$ integrated:
\begin{align}
\int d\phi_q W^{\mu\nu}
=&\frac{x_B\de(1-z_1-z_2)}{\pi x Q^2 M_p(N_c^2-1)}\int d\hat{b}_\perp \hat{b}_\perp \tilde{S}(b_\perp^2)J_0(\hat{b}_\perp \hat{q}_\perp)\left\{
A_{i}^{\mu\nu}H_{i1}\tilde{G}(x,b_\perp^2)\tilde{S}_Q^{(0)}-A_{i}^{\mu\nu}H_{i2}\tilde{H}^{\perp(2)}(x,b_\perp^2)\tilde{S}_Q^{(2)}\right\},\no
\int d\phi_q \cos(2\phi_q)W^{\mu\nu}
=&-\frac{x_B\de(1-z_1-z_2)}{\pi x Q^2 M_p(N_c^2-1)}\int d\hat{b}_\perp \hat{b}_\perp \tilde{S}(b_\perp^2)J_2(\hat{b}_\perp \hat{q}_\perp)\left\{
A_{i}^{\mu\nu}H_{i1}\tilde{G}(x,b_\perp^2)\tilde{S}_Q^{(2)}\right.\no
&\left.-A_{i}^{\mu\nu}H_{i2}\tilde{H}^{\perp(2)}(x,b_\perp^2)\frac{\tilde{S}_Q^{(0)}+4\tilde{S}_Q^{(4)}}{2}\right\},\no
\int d\phi_q \sin(2\phi_q)W^{\mu\nu}
=&\frac{x_B\de(1-z_1-z_2)}{\pi x Q^2 M_p(N_c^2-1)}\int d\hat{b}_\perp \hat{b}_\perp \tilde{S}(b_\perp^2)J_2(\hat{b}_\perp \hat{q}_\perp)\left\{
A_{i}^{\mu\nu}H_{i3}\tilde{H}^{\perp(2)}(x,b_\perp^2)\frac{\tilde{S}_Q^{(0)}-4\tilde{S}_Q^{(4)}}{2}\right\},
\end{align}
where $\hat{b}_\perp=|b_\perp|=\sqrt{-b_\perp^2}$, $\hat{q}_\perp=|q_\perp|=\sqrt{-q_\perp^2}$, and $J_0,J_2$ are $0$-th and $2$-th Bessel functions. $A_i^{\mu\nu}$ are the basic tensors we introduced in the decomposition of leptonic tensor, i.e., eq.(\ref{eq:Ai}). For simplicity, the quantity concerning $H^\perp$ is reorganized into
\begin{align}
\tilde{H}^{\perp(2)}(x,b_\perp^2)=& \frac{-2b_\perp^2}{M_p^2}\frac{\partial^2}{\partial (b_\perp^2)^2}\tilde{H}^\perp(x,b_\perp^2).
\end{align}
Due to the integration over $\phi_q$, $\phi_b$ now can be integrated. As a result, three integrated soft factors are involved, i.e.,
\begin{align}
\tilde{S}_Q^{(0)}=& \int_0^{2\pi} d\phi_b \tilde{S}^Q(b_\perp),\  \tilde{S}_Q^{(2)}= \int_0^{2\pi} d\phi_b \cos(2\phi_b) \tilde{S}^Q(b_\perp),\
\tilde{S}_Q^{(4)}= \int_0^{2\pi} d\phi_b \cos(4\phi_b)\tilde{S}^Q(b_\perp).
\end{align}
Generally, the quantities weighted by $\cos(n\phi_q)$, $\sin(n\phi_q)$ can be constructed, but more $\phi_b$ integrated soft factors will be introduced. This may not help to extract $H^\perp$.

From these weighted hadronic tensors, the corresponding cross sections can be obtained. Define weighted cross section as
\begin{align}
 \sig\langle \cos n\phi_q \rangle=\int_0^{2\pi} d\phi_q \cos(n\phi_q)\frac{d\sig}{d\psi_l dy dx_B dy_1 dy_2 d^2K_T d^2R_T}.
\end{align}
We have
\begin{align}
\sig\langle 1 \rangle=&\frac{yQ_q^2\al_{em}^2 x_B\de(1-z_1-z_2)}{64\pi^4 x Q^4 M_p(N_c^2-1)}
\int d\hat{b}_\perp \hat{b}_\perp \tilde{S}J_0(\hat{b}_\perp\hat{q}_\perp)\left\{\right.\no
&\tilde{G}\tilde{S}_Q^{(0)}\left[ H_{11}\frac{4(1-y)}{y^2}+H_{41}\frac{2(1+(1-y)^2)}{y^2}-H_{21}\frac{4\sqrt{1-y}(y-2)}{y^2}\cos\phi_l+H_{51}\frac{4(1-y)}{y^2}\cos(2\phi_l)\right]\no
&\left.-\tilde{H}^{\perp(2)}\tilde{S}_Q^{(2)}\left[ H_{12}\frac{4(1-y)}{y^2}+H_{42}\frac{2(1+(1-y)^2)}{y^2}-H_{22}\frac{4\sqrt{1-y}(y-2)}{y^2}\cos\phi_l+H_{52}\frac{4(1-y)}{y^2}\cos(2\phi_l)\right]\right\},
\end{align}
\begin{align}
\sig\langle \cos(2\phi_q) \rangle=&-\frac{yQ_q^2\al_{em}^2 x_B\de(1-z_1-z_2)}{64\pi^4 x Q^4 M_p(N_c^2-1)}
\int d\hat{b}_\perp \hat{b}_\perp \tilde{S}J_2(\hat{b}_\perp\hat{q}_\perp)\left\{\right.\no
&\tilde{G}\tilde{S}_Q^{(2)}\left[ H_{11}\frac{4(1-y)}{y^2}+H_{41}\frac{2(1+(1-y)^2)}{y^2}-H_{21}\frac{4\sqrt{1-y}(y-2)}{y^2}\cos\phi_l+H_{51}\frac{4(1-y)}{y^2}\cos(2\phi_l)\right]\no
&\left.-\tilde{H}^{\perp(2)}\frac{\tilde{S}_Q^{(0)}+\tilde{S}_Q^{(4)}}{2}\left[ H_{12}\frac{4(1-y)}{y^2}+H_{42}\frac{2(1+(1-y)^2)}{y^2}-H_{22}\frac{4\sqrt{1-y}(y-2)}{y^2}\cos\phi_l\right.\right.\no
&\left.\left.+H_{52}\frac{4(1-y)}{y^2}\cos(2\phi_l)\right]\right\},
\end{align}
\begin{align}
\sig\langle \sin(2\phi_q) \rangle=\frac{yQ_q^2\al_{em}^2 x_B\de(1-z_1-z_2)}{64\pi^4 x Q^4 M_p(N_c^2-1)}
&\int d\hat{b}_\perp \hat{b}_\perp \tilde{S}J_2(\hat{b}_\perp\hat{q}_\perp)\left\{\right.\no
&\left.\tilde{H}^{\perp(2)}\frac{\tilde{S}_Q^{(0)}-\tilde{S}_Q^{(4)}}{2}\left[ -H_{33}\frac{4\sqrt{1-y}(y-2)}{y^2}\sin\phi_l+H_{63}\frac{4(1-y)}{y^2}\sin(2\phi_l)\right]\right\}.
\end{align}
Here $H^{ij}=H^{(0)ij}+H^{(1)ij}_{finite}$, which are given in eq.(\ref{eq:hard-coeff-tree},\ref{eq:hard-part-projected}). These three weighted cross sections are our main results. We can see clearly that
$\sig\langle 1\rangle$ also depends on $H^\perp$ even if lepton azimuthal angle $\phi_l$ is integrated over, due to nonzero $\tilde{S}_Q^{(2)}$. This feature is unexpected from tree level TMD formula eq.(\ref{eq:hadronic-tree}). In addition, the soft factor $\tilde{S}_Q$ is process independent and can be absorbed into gluon distribution $\tilde{G}$ or
$\tilde{H}^\perp$, which results in so called subtracted TMDPDF\cite{Ji:2004wu}. But $\tilde{S}_Q$ is process dependent and cannot be eliminated.
Obviously, $\tilde{S}_Q$ is similar to fragmentation functions in SIDIS.  The difference is fragmentation functions can be extracted from $e^+e^-$ experiment, but $\tilde{S}^Q$ cannot be extracted in this way, because it contains a gauge link related to initial gluon. The detailed knowledge of $\tilde{S}_Q$ is beyond the scope of this paper, and we will try to study its effect on resummation in future.

\section{Summary}
In this paper we first derive the angular distributions of heavy quark pair back-to-back production in SIDIS based on tree level TMD formula. Then we examine the one-loop correction to this formula. At one-loop level a special soft factor for final heavy quarks should be complemented. From our previous studies we have known that real correction does not contribute to higher order hard coefficients. Therefore, we calculate only virtual corrections to the hadronic tensor and various nonperturbative quantities in factorization formula. Really we find the IR divergence in hadronic tensor can be absorbed by these nonperturbative quantities. As a result, we give explicit form of finite hard part, including renormalization scale dependence. Interestingly we find the IR divergent part of one-loop amplitude for heavy quark pair production is proportional to the tree level amplitude and can be expressed through standard triangle and bubble loop integrals. This feature ensures the subtraction in polarized scatterings can also be done in the same formalism. In order to present the azimuthal angle dependence about virtual photon, i.e., $\phi_q$ in hadron frame, we project the hard part into ten scalar hard coefficients. However, at one-loop level, the appearance of soft factor $S^Q$ affects the azimuthal angle dependence in a nonperturbative way, which makes the explicit $\phi_q$ dependence at cross section level become unclear. In order to extract some information about gluon TMDPDFs, especially about linearly polarized gluon distribution $H^\perp$, we construct three $\phi_q$-weighted cross sections $\sig\langle 1\rangle$, $\sig\langle\cos 2\phi_q\rangle$, $\sig\langle \sin 2\phi_q\rangle$, which depend on only three integrated heavy quark soft factors $\tilde{S}_Q^{(0),(2),(4)}$. We expect future experiments such as EIC\cite{Boer:2016fqd} can give some constraints on these three soft factors and gluon distribution $H^\perp$. For TMD factorization, an important issue is the resummation of relative transverse momentum of heavy quark pair in this process. But due to the many hard scales in this process, e.g., $Q^2, R_\perp^2$, and the scales introduced by soft factors like $\zeta^2$, $4v_1\cdot\tilde{v}v_2\cdot\tilde{v}/\tilde{v}^2$, the resummation will be nontrivial. We want to study this issue in another paper. In appendix we present all involved IR finite loop integrals, which are used to express the finite hard part $H_2^{ij}$.

\section*{Acknowledgements}
The author would like to thank Jian-Ping Ma for reading the manuscript. The hospitality of ITP, CAS during the completion of this paper is appreciated. This work is supported by National Nature Science Foundation of China(NSFC) with contract No.11605195.

\appendix
\section{IR finite scalar integrals}
Following scalar integrals are used to express the finite result. These integrals can be obtained from general results in \cite{Denner:1991kt} or references therein by proper continuation to DIS region. Straightforward calculation is also easy. Most of the results appear very simple. All of these functions may depend on $s_1$, therefore $s_1$ is suppressed in the arguments. $\tx{DBox1}$, $\tx{DBox2}$ have been defined in eq.(\ref{eq:DBox1},\ref{eq:DBox2}). Their explicit expressions are
\begin{align}
\tx{DBox1}[t_1,u_1]=&\frac{-i}{16\pi^2}\frac{1}{s_{12}u_1\rho_{12}}\left[
-2\ln c_{12}\ln\frac{r_b}{-y_{23}}+2\ln(-c)\ln\frac{\rho-\rho_{12}}{\rho+\rho_{12}}
-\ln^2(-c)+2\ln\frac{\rho^2-1}{\rho^2-\rho_{12}^2}\ln c_{12}\right.\no
&\left.+4Li_2(-c_{12})-2Li_2(\frac{c_{12}}{c})-2Li_2(cc_{12})
\right],\no
\tx{DBox2}[t_1u_1,t_1+u_1]=&\frac{-i}{16\pi^2}\frac{(Q^2)^{-2}}{y_{13}y_{23}}\left[\frac{2}{3}\pi^2
+\ln^2(-c)+\ln^2\frac{y_{13}}{y_{23}}
+Li_2(\frac{r_b+y_{13}}{r_b})+Li_2(\frac{r_b+y_{23}}{r_b})\right],
\end{align}
where $\rho=\sqrt{1-4m^2/q^2}>1$, $\rho_{12}=\sqrt{1-4m^2/s_{12}}$ and $c=(1-\rho)/(1+\rho)$, $c_{12}=(1-\rho_{12})/(1+\rho_{12})$.

IR finite triangle integrals are
\begin{align}
\tx{CTri3}(u_1)=&\int_l\frac{1}{[l^2+i\ep][(l-k_2)^2-m^2+i\ep][(l-k_2+p)^2-m^2+i\ep]},\no
\tx{CTri4}(t_1,u_1)=&\int_l\frac{1}{[l^2+i\ep][(l+p-k_2)^2-m^2+i\ep][(l+k_1)^2-m^2+i\ep]},\no
\tx{CTri5}(t_1 u_1, t_1+u_1)
=&\int_l\frac{1}{[l^2-m^2+i\ep][(l+p)^2-m^2+i\ep][(l+p-k_1-k_2)^2-m^2+i\ep]}.
\end{align}
In DIS region, the results of $\tx{CTri3}$ and $\tx{CTri5}$ appear very simple
\begin{align}
\tx{CTri3}(u_1)=& -\frac{i}{16\pi^2}\frac{1}{u_1}
\left\{-\frac{\pi^2}{6}+Li_2(\frac{r_b+y_{23}}{r_b})\right\},\no
\tx{CTri5}(t_1 u_1, t_1+u_1)=& \frac{i}{16\pi^2}\frac{1}{2(t_1+u_1)}
\left\{\pi^2+ \ln^2(-c)-\ln^2 c_{12}\right\}.
\end{align}
But that for $\tx{CTri4}$ is a little complicated
\begin{align}
\tx{CTri4}(t_1,u_1)=&\frac{i}{16\pi^2}\frac{1}{Q^2(\be_1-\be_2)}
\left\{ K(\be_1,\be_3)+K(\be_1,\be_4)-K(\be_2,\be_3)-K(\be_2,\be_4)\right.\no
&\left.+\frac{\pi^2}{2}+\frac{1}{2}\ln^2\be_1 - \frac{1}{2}\ln^2(-\be_2)
-\ln\frac{-u_1}{Q^2}\ln\frac{-\be_2(1-\be_1)}{\be_1(1-\be_2)}
+Li_2(\be_1)-Li_2(\be_2)\right\},
\end{align}
where
\begin{align}
\be_{1}=\frac{1+y_{23}+\sqrt{(1+y_{23})^2+4r_b}}{2},\ \be_{2}=\frac{1+y_{23}-\sqrt{(1+y_{23})^2+4r_b}}{2},
\end{align}
and
\begin{align}
\be_{3}=\frac{1+\sqrt{1+4r_b}}{2},\ \be_{4}=\frac{1-\sqrt{1+4r_b}}{2}.
\end{align}
In DIS region, one can show $\be_2<\be_4<0<\be_1<1<\be_3$. What is crucial is $\be_1<1$, which ensures the functions in the result are real. That is,
\begin{align}
K(\be_1,\be_3)\doteq& \ln(\be_3-1)\ln\frac{1-\be_1}{\be_3-\be_1}-
\ln(\be_3-\be_1)\ln\frac{\be_1}{\be_3-\be_1}-\frac{7\pi^2}{6}
+Li_2(\frac{1-\be_3}{\be_1-\be_3})+Li_2(\frac{\be_1}{\be_1-\be_3}),\no
K(\be_1,\be_4)\doteq& \ln(\be_1-\be_4)\ln\frac{1-\be_1}{\be_1-\be_4}-
\ln(-\be_4)\ln\frac{\be_1}{\be_1-\be_4}+\frac{\pi^2}{6}
-Li_2(\frac{\be_1-1}{\be_1-\be_4})-Li_2(\frac{-\be_4}{\be_1-\be_4}),\no
K(\be_2,\be_3)\doteq& \ln(\be_3-1)\ln\frac{1-\be_1}{\be_3-\be_2}-
\ln\be_3\ln\frac{-\be_2}{\be_3-\be_2}
+Li_2(\frac{\be_3-1}{\be_3-\be_2})-Li_2(\frac{\be_3}{\be_3-\be_2}),\no
K(\be_2,\be_4)\doteq& \ln(1-\be_4)\ln\frac{1-\be_2}{\be_4-\be_2}-
\ln(-\be_4)\ln\frac{-\be_2}{\be_4-\be_2}
+Li_2(\frac{1-\be_4}{\be_2-\be_4})-Li_2(\frac{-\be_2}{\be_2-\be_4}),
\end{align}
where all imaginary parts are dropped.

In addition, bubbles $B_0(\tilde{p}^2,m_1^2,m_2^2)$ also appear in final result. They are UV divergent and thus $\mu$ dependent. According to our previous classification, the UV divergence in $B_0$ functions are removed by $\overline{MS}$-scheme, and the $\mu$ dependence or factor $\ln\mu/Q$ is absorbed into $H_3$. Thus, following finite $b_0$ functions are used to express $H_2$:
\begin{align}
b_0(\tilde{p}^2,m_1^2,m_2^2)=B_0(\tilde{p}^2,m_1^2,m_2^2)
-\frac{i}{16\pi^2}\left(\frac{2}{\ep}-\ga_E+\ln\frac{4\pi\mu^2}{Q^2}\right).
\end{align}
where
\begin{align}
B_0(\tilde{p}^2,m_1^2,m_2^2)=&\int_l\frac{1}{[l^2-m_1^2+i\ep][(l+\tilde{p})^2-m_2^2+i\ep]}.
\end{align}
Besides, there are two special bubble integrals $B_v, B_m$ by mass renormalization for Fig.\ref{fig:oneloop}(j,k). They have been given in eq.(\ref{eq:BvBm}).

For completeness we also present the explicit expressions of $b_0(\tilde{p}^2,m_1^2,m_2^2)$ functions here. In our case, the involved $b_0$ functions can be divided into two classes: $m_1=m_2=m$ and $m_1=0$ or $m_2=0$. For the first class we have
\begin{align}
b_0(\tilde{p}^2,m^2,m^2)=&\frac{i}{16\pi^2}\left(
2+\ln\frac{Q^2}{m^2}+\omega\ln\frac{\omega-1}{\omega+1}\right),\hspace{0.5cm}
\omega=\sqrt{1-\frac{4m^2}{\tilde{p}^2+i\ep}}.
\end{align}
Two special cases are involved in the result: $\tilde{p}^2=s_1+2m^2$ and $\tilde{p}^2=q^2=-Q^2$. After continuation the real parts of these two $b_0$ functions are
\begin{align}
b_0(s_1+2m^2,m^2,m^2)=&\frac{i}{16\pi^2}\left(
2+\ln\frac{Q^2}{m^2}+\omega\ln\frac{1-\omega}{1+\omega}\right),\hspace{0.5cm}
\omega=\sqrt{1-\frac{4m^2}{s_1+2m^2}},\no
b_0(-Q^2,m^2,m^2)=&\frac{i}{16\pi^2}\left(
2+\ln\frac{Q^2}{m^2}+\omega\ln\frac{\omega-1}{\omega+1}\right),\hspace{0.5cm}
\omega=\sqrt{1+\frac{4m^2}{Q^2}}.
\end{align}
For the second class of $b_0$ functions, one inner propagator is massless. The general result is
\begin{align}
b_0(\tilde{p}^2,0,m^2)=&\frac{i}{16\pi^2}\left(
2+\ln\frac{Q^2}{-\tilde{p}^2+m^2}+\frac{m^2}{\tilde{p}^2}
\ln\frac{-\tilde{p}^2+m^2}{m^2}\right).
\end{align}
There are three special cases in our calculation: $1) \tilde{p}^2=0$, $2)\tilde{p}^2=m^2$, $3)\tilde{p}^2<0$. By taking proper limit the first two cases can be obtained
\begin{align}
b_0(0,0,m^2)=&\frac{i}{16\pi^2}\left(1+\ln\frac{Q^2}{m^2}\right),\no
b_0(m^2,0,m^2)=&\frac{i}{16\pi^2}\left(2+\ln\frac{Q^2}{m^2}\right).
\end{align}
The last case is well-defined from the general expression. Specifically, $\tilde{p}^2=m^2+t_1$ or $m^2+u_1$ in our calculation, and both are negative in DIS region.

\bibliography{ref2}

\end{document}